\documentclass[aps,pra,twocolumn,showpacs,superscriptaddress, nofootinbib]{revtex4-2}   
\usepackage{graphicx}  
\usepackage{epstopdf}
\usepackage[caption=false]{subfig}
\usepackage{float} 
\usepackage[toc,page]{appendix}
\usepackage{dcolumn}   
\usepackage{bm}        
\usepackage{amssymb}   
\usepackage{mathrsfs}   
\usepackage{color}
\usepackage{hyperref}
\usepackage{mathtools}
\usepackage{multirow}
\usepackage[table]{xcolor}
\usepackage{amsmath, amsthm, amssymb}
\usepackage{amsfonts}
\usepackage{bbold}       
\usepackage{accents}    

\hypersetup{
    colorlinks=true, 
    linktoc=all,     
    linkcolor=blue,  
    citecolor=blue,
    filecolor=blue,
    urlcolor=blue
}

\usepackage{soul}
\usepackage{cancel}
\usepackage{cleveref}

\usepackage{xcolor}
\pagecolor{white}

\newcommand{\beq}{\begin{equation}}
\newcommand{\eeq}{\end{equation}}
\newcommand{\bqa}{\begin{eqnarray}}
\newcommand{\eqa}{\end{eqnarray}}
\newcommand{\bse}{\begin{subequations}}
\newcommand{\ese}{\end{subequations}}
\newcommand{\nn}{\nonumber}

\newcommand{\erf}[1]{Eq.~(\ref{#1})}

\newcommand{\erfa}[2]{Eqs.~(\ref{#1}) and (\ref{#2})}
\newcommand{\arf}[1]{Appendix~\ref{#1}} 
 
\newcommand{\srf}[1]{Sec.~\ref{#1}} 
\newcommand{\crf}[1]{Ref.~\cite{#1}}

\newcommand{\frf}[1]{Fig.~\ref{#1}}

\newcommand{\eg}{{\it e.g.~}}
\newcommand{\dg}{^\dagger}
\newcommand{\expt}[1]{\langle{#1}\rangle}

\definecolor{BLACK}{gray}{0}
\definecolor{dgray}{rgb}{0.26,0.26,0.26}
\definecolor{RED}{rgb}{1,0,0}
\definecolor{GREEN}{rgb}{0.2,.5,0.2}
\definecolor{BLUE}{rgb}{0,0,1}
\definecolor{AWESOME}{rgb}{0.54,0.17,0.29}
\definecolor{BEN}{rgb}{0.3,0.4,0.6}
\definecolor{ALEXEI}{rgb}{0,0.6,0.3}

\newcommand{\blk}{\color{BLACK}}

\renewcommand{\(}{\left(}
\renewcommand{\)}{\right)}
\newcommand{\sq}[1]{\left[{#1}\right]}
\newcommand{\cu}[1]{\left\{ {#1} \right\}}

\newcommand{\abs}[1]{\left| {#1} \right|}
\newcommand{\tr}[1]{{\rm Tr}\sq{ {#1} }}

\newcommand{\smallfrac}[2]{\mbox{$\frac{#1}{#2}$}}

\newcommand{\bra}[1]{\langle{#1}|}
\newcommand{\ket}[1]{|{#1}\rangle}
\newcommand{\ip}[2]{\langle{#1}|{#2}\rangle} 
\newcommand{\op}[2]{\ket{#1}\bra{#2}}
\newcommand{\s}[1]{ \hat{\sigma}_{#1}}

\newcommand{\gc}{ {\rm g} }
\newcommand{\er}{ {\rm e} }

\newcommand{\vo}{\hat{\varrho}}

\newcommand{\g}[1]{\gamma_{#1}}

\newcommand{\typea}{\Phi} 
\newcommand{\typeb}{\Psi} 

\newcommand{\Ldiff}{\hat{L}'} 
\newcommand{\Ldiffdg}{\hat{L}'\phantom{}^\dagger} 

\newcommand{\typeasuperscript}{}
\newcommand{\typeasuperscriptdag}{^\dagger}

\begin{document}

\widetext


\title{Collisional-model quantum trajectories for entangled qubit environments}

\author{Shakib Daryanoosh} \email{shakib.daryanoosh@uwa.edu.au}
\affiliation{Centre for Engineered Quantum Systems, Department of Physics and Astronomy, \\ Macquarie University, Sydney, NSW 2122, Australia}
\affiliation{Department of Physics, University of Oxford, Clarendon Laboratory, Parks Road, Oxford OX1 3PU, UK}
\affiliation{Department of Physics, University of Western Australia, Crawley, WA 6009, Australia}
\author{Alexei Gilchrist}
\affiliation{Centre for Engineered Quantum Systems, Department of Physics and Astronomy, \\ Macquarie University, Sydney, NSW 2122, Australia}
\author{Ben Q. Baragiola} 
\affiliation{Yukawa Institute for Theoretical Physics, Kyoto University, Kitashirakawa Oiwakecho, Sakyo-ku, Kyoto 606-8502, Japan}
\affiliation{Centre for Engineered Quantum Systems, Department of Physics and Astronomy, \\ Macquarie University, Sydney, NSW 2122, Australia}
\affiliation{Centre for Quantum Computation and Communication Technology, School of Science, \\ RMIT University, Melbourne, Victoria 3001, Australia}

%
%


%
\vskip 0.25cm

\date{\today}

\begin{abstract} 
We study the dynamics of quantum systems interacting with a stream of entangled qubits. Under fairly general conditions, we present a detailed framework describing the conditional dynamical maps for the system, called quantum trajectories, when the qubits are measured. 
Depending on the measurement basis, these quantum trajectories can be jump-type or diffusive-type, and they can exhibit features not present with quantum optical and single-qubit trajectories. 
As an example, we consider the case of two remote atoms, where jump-type quantum trajectories herald the birth and death of entanglement.
\end{abstract}

\maketitle

\section{Introduction}

The study of open quantum systems is important, because no quantum system is fully isolated from its environment. Mitigating environmental decoherence is crucial for advancement of technologies that exploit intrinsic quantum properties---coherent superposition and entanglement---including quantum computing, communication, and metrology~\cite{BarKim17}. 
Although leakage from a system into its environment is inevitable, measurement of the environment itself restores lost information. The recovered information, in the form of random measurement outcomes, can be used to provide an update of the system's state given by a series of stochastic maps generically called a \emph{quantum trajectory}~\cite{Molmer92,Carmichael93,Bru02,Siddiqi14,Daley14,Huard16}. 

Much effort has gone into the treatment of bosonic-field environments in the quantum optics community~\cite{Gardiner85, Car08, BrePet03, WisMil10} due to the fact that electromagnetic modes are ubiquitous in nature.
Recent attention has focused on an alternative~\cite{2021campbell60001}: repeated interaction models, also known as collisional models, where the environment (referred to henceforth as a \emph{bath}) is a series of discrete quantum systems---typically qubits---each of which interacts with the system of interest for a short time and then propagates away~\cite{NogHar99}. Unconditional dynamics for the system are found by tracing out the bath qubits, and measuring each qubit after the interaction generates quantum trajectories~\cite{AttJoy07, Pel10b, BruMer14}. 
Collisional models have gained prominence for their relative simplicity, for applications to quantum thermodynamics~\cite{Hor12, StrEsp17}, and for investigating future technologies such quantum-limited thermometry~\cite{Seah19} and rapid charging of quantum batteries~\cite{Seah21}.
Futher, collisional models can simulate bosonic~\cite{GroCom17} and other environments~\cite{GioPal12,CatGio21}.

Collisional models typically employ single-qubit baths, where a single, uncorrelated qubit interacts with the system at each time, with recent investigations pushing beyond this in various ways. Temporally correlated single-qubit baths can be used to model non-classical environments such as $n$-photon states \cite{DabChr17,Dabrowska2019,Luch22}, which drive non-Markovian evolution due to the temporal entanglement in the input bath~\cite{Dabrowska21}. Another way to to go beyond single-qubit baths is to increase the number of qubits per interaction time. Two-qubit collisional models were found to have the power to entangle separated quantum systems and to generate high levels of two-mode squeezing between remote cavities~\cite{DarGil18}.

In this work, we build on the average, master equation analysis in Ref.~\cite{DarGil18} and develop a general framework for quantum trajectories for two-qubit baths. This framework extends beyond prior single-qubit trajectory work to situations where coherences/entanglement among the bath qubits can be converted into quantum correlations across the systems \cite{WuGu17, CerSou17, DarGil18}. This happens stochastically and is highly dependent on the measurement basis, which can be chosen to mimic the two standard types of trajectories in quantum optics: jump-type and diffusive-type. Moreover, collisional models can go beyond input-output quantum optical models, which are bound by physical constraints that limit the excitation probability in the bath over small time intervals~\cite{Mollow68, WisMil10, Baragiola:2017aa}. We identify a new type of conditional dynamics arising from this fact that we call \emph{quasi-diffusive}, which has features of diffusion but produces less pronounced effects.

This paper is organized as the following. In \srf{sec:physmod} we lay out the physical model and present necessary background on dynamical maps and quantum trajectories. 
To set the stage for the quantum trajectories, in \srf{sec:MEsection}, we review the two-qubit bath states and corresponding master equations from Ref.~\cite{DarGil18}.
Section~\ref{sec:quantumtrajectories} contains the main results of this work: quantum trajectories for two-qubit baths. Focusing on local measurements of the bath qubits in their energy eigenbases, we derive the Kraus operators, conditional maps, outcome probabilities, and conditional difference equations. These form a toolbox with which one can construct quantum trajectories for other bases, including entangled Bell-basis measurements and local mixed measuremends, where one of the two qubits is measured in a basis orthogonal to its energy basis.    
In \srf{sec:eng:base}, we show that by coarse graining over some of the outcomes in the quantum trajectories, one can derive standard jump-type and diffusive-type stochastic master equations (SMEs).
Section~\ref{2qubitsection} illustrates features of two-qubit trajectories with an example, where the system itself is comprised of a pair of remote, two-level atoms. 
In Sec.~\ref{sec:beyond2qubits}, we discuss the consequences of considering more than two qubits in the bath within the weak-coupling limit, and we illustrate them with a three-qubit example. 
Finally, in \srf{sec:con}, we summarize the findings of this paper and consider avenues for further study.

\section{Physical model} \label{sec:physmod}
To describe the underlying framework behind our conceptual model, we begin with the concept collisional models (or repeated quantum interactions)~\cite{AttPau06, AttJoy07, BruMer14, StrEsp17, GroCom17}. Within this formalism, a quantum system in the Hilbert space ${\mathscr H}_{S}$ couples to an bath comprised of a stream of identical and independent quantum systems ${\mathscr H}_{B}^{(k)}$, such that ${\mathscr H}_{B} = \bigotimes_k {\mathscr H}_{B}^{(k)}$. 
This stream sequentially couples to the system, with each element interacting over a short time interval of duration $\Delta t \coloneqq t_{k+1} - t_{k}$, after which it can (a) carry on unimpeded or (b) undergo a projective measurement. If it is measured, the process is equivalent to a weak measurement of the system, resulting in conditional evolution in ${\mathscr H}_{S}$~\cite{WisMil10}. 

A similar scenario arises in quantum optical settings, where a probe field interacts with the system continuously in time. However, the situation here is different, because system-bath coupling is discrete, as it is packaged into time steps of duration $\Delta t$. 
This is not a critical distinction, as the transition from discrete repeated interactions to continuous-in-time is well defined~\cite{AttPau06, AttJoy07, Pel08, Pel10a, Pel10b}. More importantly, the bath is made up of two-level systems (qubits) rather than bosonic modes. 

In each time interval $\Delta t$ the environment is composed of $n$ potentially entangled qubits, each of which couples to its own subsystem. In this work we will focus on $n=2$ as this is sufficient to capture interesting features, see \frf{fig:setup}. It is important to note that while the bath qubits may be entangled within each interaction time interval, they are not entangled \emph{between} time intervals. This type of quantum correlated environment underlies non-Markovian dynamics \cite{Korotkov:2002aa,DabChr17,Baragiola:2017aa} which should be treated differently~\cite{CicPal21,Luch22}.

\begin{figure}
	\includegraphics[scale=0.15]{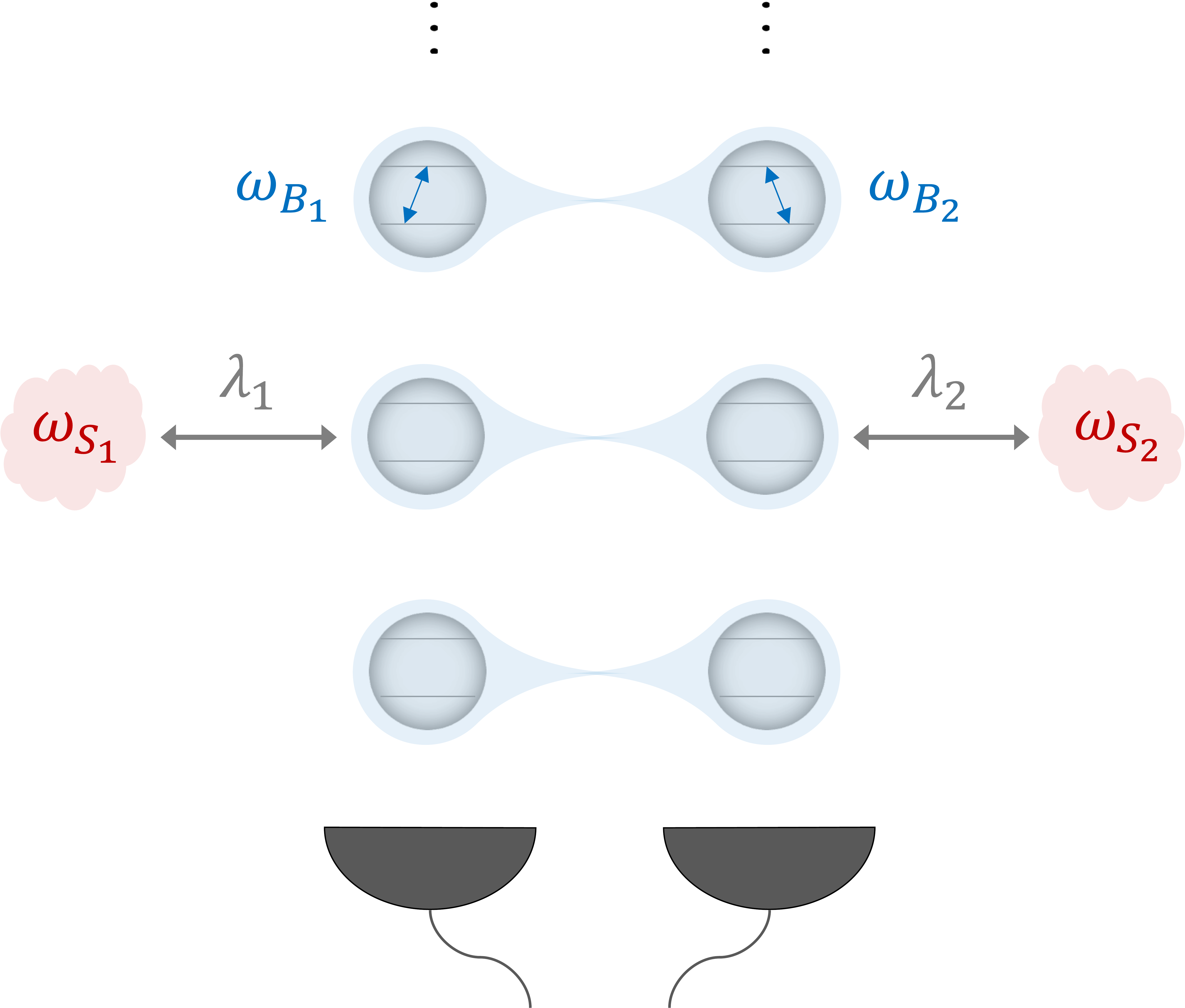}
	\caption{\label{fig:setup} (Color online). Conceptual diagram of the physical model. A stream of entangled two-level quantum systems (proceeding from top to bottom here) interact, one after another, with two separate subsystems. After each interaction, the qubits are projectively measured, and the outcomes are used to update the joint state of the two subsystems.
	}
\end{figure}

For each interaction time step the total Hilbert space of the system plus environment is ${\mathscr H} = {\mathscr H}_{S} \otimes {\mathscr H}_{B}^{(k)}$, and the evolution of the joint system-environment is generated by the following Hamiltonian
\beq
\hat H = \hat H_S + \hat H_B^{(k)} + \hat H_{SB}^{(k)},
\eeq
where $\hat H_S$ and $\hat H_B^{(k)}$ are the Hamiltonians of the system and environment respectively, and $\hat H_{SB}^{(k)}$ is the system-environment interaction Hamiltonian, expressed as the following
\bqa
{ \hat H_S} &{=}& { \sum_{\ell=1,2}  \omega_{\textit{S}_\ell} \hat c_{\ell}\dg \hat c_{\ell},  \quad \hat H_B^{(k)} = \sum_{\ell=1,2} \omega_{\textit{B}_\ell} \s{\ell}\dg \s{\ell},} \\
\hat{H}^{({k})}_{SB} &=& \sum_{\ell=1,2} \lambda_\ell \big( \hat c_\ell\dg \s{\ell} + \hat c_\ell \s{\ell}\dg \big).
\eqa
Here $\omega_{\textit{S}_\ell}$ and $\omega_{\textit{B}_\ell}$ are the characteristic frequencies of the subsystems and qubit bath, respectively; $\lambda_\ell$ is the coupling strength between the $\ell$-th subsystem and its qubit; the system operators $\hat{c}_\ell$ are eigenoperators of the system Hamiltonian~\cite{Bar15}; $\s{\ell} \coloneqq \ket{\gc}_\ell\bra{\er}$ is a lowering operator acting on the space of the qubit bath; and $\s{\ell}\dg$ is the corresponding raising operator. Since the bare Hamiltonian $\hat H_S + \hat H_B^{(k)}$ generates well understood free evolution, we move to the interaction picture in order to highlight the interesting dynamics. In this picture the interaction Hamiltonian becomes
\beq  \label{IPIH:2q}
\hat{H}^{({k})}_{SB} \rightarrow \sum_{\ell=1,2} \lambda_\ell \big( \hat c_\ell\dg \s{\ell} e^{-i \delta_\ell t} + {\rm H. c.} \big),
\eeq 
where $\delta_\ell \coloneqq \omega_{\textit{B}_\ell} - \omega_{\textit{{S}}_\ell}$ is the detuning and H.c. is the Hermitian conjugate of the term. 

In this work, we specifically investigate on-resonant system-environment couplings ($\delta_\ell=0$) for two qubits as this is sufficient to capture interesting behaviour.\footnote{If multiple coupling rates $\lambda_\ell$ vary considerably, the respective qubits should be considered as separate baths with different effective rates and separate Lindblad operators in a master-equation setting. Here, we aim to consider the grouping of two qubits as a single bath so $\lambda_1 \approx \lambda_2$.} We work in the weak-coupling regime, where system-bath interactions, given by \erf{IPIH:2q}, generate only weak correlations between each pair of bath qubits and their respective systems. Weak coupling occurs in settings where the coupling rates $\lambda_\ell$ are small or where the interactions are fleeting (small $\Delta t$). 
In this limit, we can expand the unitary evolution operator for the joint system-environment, $\hat{U}_{SB}^{(k)}=\exp\{-i\hat{H}_{SB}^{({k})}\Delta t\}$, to second order in $\Delta t$~\cite{CicPal21}:
\begin{align}  \label{Utaylor}
\hat{U}_{SB}^{({k})} 
	& = \hat{I}_{SB} - i \hat{H}^{({k})}_{SB} \Delta t - \tfrac{1}{2} \big(\hat{H}^{({k})}_{SB} \big)^2 {\Delta t}^2 + \mathcal O({\Delta t}^3). 
\end{align}
Note that although $\lambda_\ell \Delta t$ is assumed small, it is nonvanishing. Indeed, the second order term must be kept to retain unitarity, $\hat{U}_{SB}^{({k})\dagger} \hat{U}_{SB}^{({k})} = \hat{I}_S \otimes \hat{I}_B +  \mathcal{O}(\Delta t^3)$. Another approach defines an effective coupling parameter 
	\begin{equation} \label{weakcouplingrate}
		\gamma_\ell \coloneqq \lambda_\ell^2 \Delta t  \, ,
	\end{equation}
interpreted as a rate such that the weak-coupling limit keeps terms to first order in $\gamma_\ell \Delta t$~\cite{vanHove1954, Accardi2002} and discards higher-order terms. 
Regardless, with either method, the result is \erf{Utaylor}.
Note that weak coupling does not mean that the interactions are continuous-in-time---that limit is taken separately~\cite{CicPal21}.

\subsection{Unconditional maps: master equations}

A dynamical map for the unconditional (\emph{i.e.} average) reduced system state is obtained by tracing over the bath after the interaction. The reduced state could describe a single quantum system or could be the joint state of multiple subsystems interacting with the same bath.  
For a single pair of bath qubits weakly interacting over a time $\Delta t$, the unconditional dynamical map ${\cal M}$, also called the ``collision map''~\cite{CicPal21},  is
\beq \label{map:gen}
\mathcal M \big[ \hat{\rho}(t_{k}) \big]\coloneqq {\rm Tr}_B \left\{ \hat{U}_{SB}^{({k})} \big[ \hat \rho{(t_{k})} \otimes \hat \rho_B^{({k})} \big] \hat{U}_{SB}^{({k})}{}\dg \right\}. 
\eeq
The joint state of the environment is assumed to be of the form 
\beq \label{initstate}
\hat \rho_{B} \equiv \bigotimes_ {k} {\hat{\rho}}_{B}^{({k})},
\eeq
where each pair of entangled bath qubits $k$ is described by $\hat \rho_B^{({k})}$ and is independent of other pairs in the  chain. This guarantees the resulting map, \eqref{map:gen}, is completely positive, trace-preserving (CPTP), and  Markovian---in fact, it is CP divisible~\cite{CicPal21}.

In an appropriate limit, the dynamical map can be described by a continuous-in-time Markovian master equation. The details for deriving such a ME can be found, for instance, in \crf{GriMar16, CicPal21, DarGil18}. In the following, for the clarity of notation we drop the explicit superscripts for the system and environment states unless confusion could arise.

\subsection{Conditional maps: quantum trajectories} \label{sec:quantumtrajectories}

After interacting with the system, the bath contains system information that can be recovered upon measurement. This recovered information can be used to construct a conditional map describing the outcome-dependent evolution of the reduced system. Since quantum measurements yield random outcomes, these conditional reduced-state maps are stochastic in nature, and the form of these maps depends on the basis in which the environmental measurement is performed.
For a given measurement record, which is in general a string of measurement results, the conditional reduced system state explores a path in the Hilbert space of the system ${\mathscr H}_S$ called a quantum trajectory \cite{Car08}. We briefly review quantum trajectories and discuss the two commonly studied examples, jump-type and diffusive-type, in repeated interaction and quantum optical settings.

Consider the canonical case of a system prepared in state $\vo$ and a single auxiliary system, which acts as the bath, prepared in the pure state $\op{\psi_B}{\psi_B}$.  
After unitary interaction via $\hat U_{SB}$, the bath is projectively measured in some basis resulting in a single outcome $m$ associated with eigenstate $\ket{m}$.\footnote{We take the measurements to be \emph{sharp}, that is, the measurement projectors are rank-1.}
The map ${\mathcal E}_m$ updates the state of the system conditioned on the measurement outcome~\cite{WisMil10},
\bse
\begin{align}   
{\mathcal E}_m(\vo) & \coloneqq \bra{m} \hat{U}_{SB} \left(\hat{\varrho} \otimes \op{\psi_B}{\psi_B}\right) \hat{U}^\dagger_{SB} \ket{m}, \\
 & = \hat K_m \vo \hat K_m\dg, \label{map:condgen}
\end{align}
\ese
where in the second line we define Kraus operators
\beq \label{Kraus:op}
\hat K_m \coloneqq \bra{m} \hat U_{SB} \ket{\psi_B}
\, ,
\eeq
which satisfy the completeness relation $\sum_m \hat K_m\dg \hat K_m = \hat{I}_S$, with the sum taken over all possible outcomes. 
The trace of the output of the map \erf{map:condgen} encodes the probability of outcome $m$,
\beq \label{prob:cond}
\wp_m = \text{Tr} \big[\hat K_m \vo \hat K_m\dg \big], 
\eeq
so that the normalised output state is  $\hat\varrho_\text{out} =  \wp_m^{-1} {\mathcal E}_m(\vo)$.

We apply the above formalism to the case of repeated weak interactions (or ``collisions'') described by \erf{Utaylor}, where the environmental state \erf{initstate} is pure and identical across all time slices, $\hat{\rho}_B^{(k)} = \op{\psi_B}{\psi_B}$ (the mixed-state case is discussed in \arf{appn:mixed:bath}). A quantum circuit representation of this situation is shown in \frf{fig:circuit}. For outcome $m$, the conditional reduced state at each time step is given by the discrete dynamical map
\beq \label{map:cond}
\vo(t_{k +1}) = \frac{1}{\wp_m}  \mathcal E_m \big[ \vo(t_{k}) \big] = \frac{1}{\wp_m} \hat K_m \vo(t_k) \hat K_m\dg, 
\eeq
with weak-measurement Kraus operators
\beq  \label{KrausWeak}
\hat K_m 
=  \ip{m}{\psi_B} - i \Delta t\bra{m} \hat{H}^{({k})}_{SB} \ket{\psi_B}  - \tfrac{1}{2} {\Delta t}^2 \bra{m} \big(\hat{H}^{({k})}_{SB} \big)^2  \ket{\psi_B} 
\, , 
\eeq
whose explicit form can be found given a specific environmental state and measurement basis.
The unconditional map in \erf{map:gen} can be obtained by taking the average over conditional maps corresponding to all measurement outcomes.

\begin{figure}
	\includegraphics[scale=0.14]{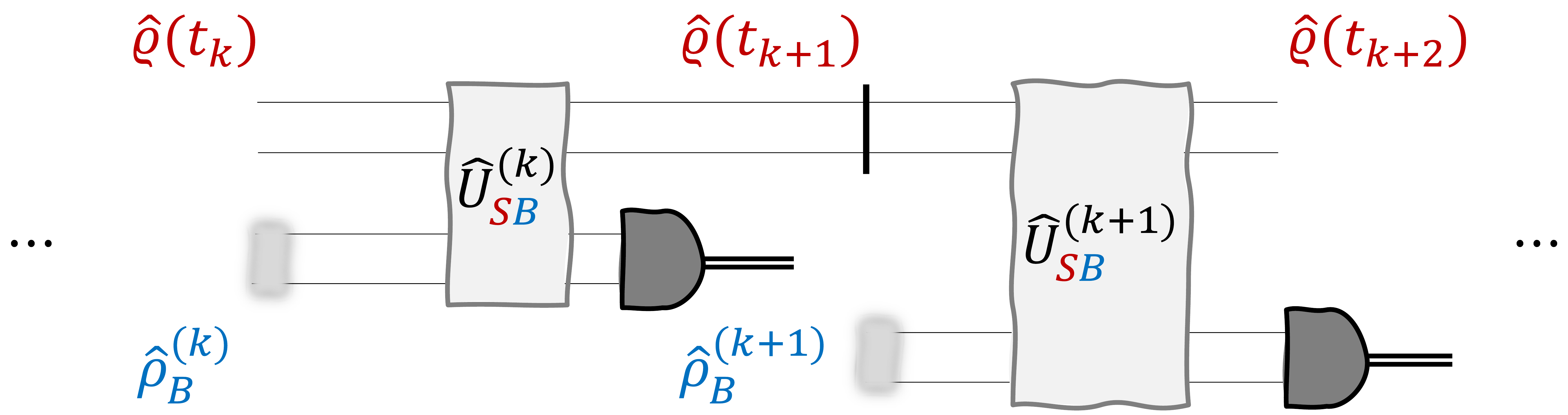}
	\caption{\label{fig:circuit} Circuit diagram illustration of quantum repeated interactions. 
	The first $(k-1)$ interactions and measurements have prepared the system state $\hat{\rho}(t_k)$ for the collision with the environmental system prepared in $\hat \rho_B^{(k)}$. They interact and become correlated via the unitary $\hat{U}_{SB}^{(k)}$, after which the environment is measured. The outcome of that measurement determines the conditional system state, $\vo(t_{k+1})$.
}
\end{figure}

Often one works with difference maps (or differential maps) rather than the state update maps. For our situation, the conditional difference map takes the form~\cite{Francesco:2017aa},
	\beq \label{dffmap}
	\Delta \hat{\varrho}_m(t_k) = \hat \varrho(t_{k+1})  - \hat \varrho(t_k) = ( \wp_m^{-1} \mathcal{E}_m - \mathcal{I}) \big[\hat{\varrho}(t_k) \big]
	\, ,
	\eeq
with $ \mathcal{I}$ being the identity map.
The transition between maps of this type arising from discrete repeated quantum interactions
and those in continuous quantum measurement theory has been established~\cite{Pel08,Pel10a,Pel10b}, wherein discrete quantum trajectories---classical Markov chains that describe the evolution of the system in the context of quantum repeated interactions---can be interpreted as solutions to discrete stochastic differential equations (SDEs). By allowing the interaction time interval to go to zero $\Delta t \rightarrow 0$ while properly scaling the interaction strength,
$\lambda_\ell \sim (\Delta t)^{-1/2}$~\cite{Francesco:2017aa,CicPal21},
the continuous-limit solutions satisfy SDEs describing continuous quantum trajectories. The mathematical tools required to show this are related to convergence of SDEs and/or general martingale problems in probability theory; we direct interested readers to \crf{Pel10b} for more information.
Practically, though, the explicit convergence of discrete quantum trajectories to continuous ones is less valuable, as any physical settings involve digitization of measurements through integration of detector currents, and numerical simulations also inevitably employ discretization.

\subsubsection{Jump-type and diffusion-type dynamics}  \label{ConditionalTypes}

Both discrete and continuous quantum trajectories have been widely studied for single-qubit baths~\cite{CicPal21}. 
\blk We briefly review two critical types of conditional evolution---``jump''-type evolution and ``diffusion''-type evolution---that arise for single-qubit baths~\cite{GroCom17} and also appear in quantum optical settings with bosonic-mode baths~\cite{Car08, WisMil10}. 
There, the bath state is typically a Gaussian state of the field, and it is the measurement basis---typified by photon counting or homodyne detection---that determines the nature of the conditional map. 
Qubit bath models can be used to emulate quantum optical cases~\cite{GroCom17,CicPal21}, and qubit baths have the additional freedom that input bath states can be more exotic~\cite{CicPal21}. 

Consider a single quantum system interacting with a stream of bath qubits.
For the model in \srf{sec:physmod}, this is described by the second system and its associated probe qubit being absent
(also, set $\lambda_1=\lambda$ and $\hat{c}_1=\hat{c}$ for convenience). 
Suppose that each bath qubit is prepared in the ground state $\ket{\gc}$ and is measured in the $\{\gc, \er \}$ basis, akin to photon counting in quantum optics. Detection of a bath qubit in the excited state $\ket{\er}$ (orthogonal to its input state) indicates a quantum jump, where the system state abruptly changes in a significant way. 
The probability of a quantum jump to take place is $\wp_\text{jump} = \gamma \expt{\hat c\dg \hat c} \Delta t = O(\gamma \Delta t)$, where $\gamma \coloneqq \lambda^2 \Delta t$ is the jump rate, and $\expt{\star}$ is the expectation value with respect to the system state. 
Since $\gamma \Delta t \ll 1$, these quantum jumps are rare. 
The probability of no jump occurring is $1 - \wp_{\rm jump} = 1 - O(\gamma \Delta t)$.
Thus, most of the time, the bath qubits are found to be in $\ket{\gc}$, and the system evolves more smoothly under a ``no-jump'' map with only infinitesimal changes to the system state. 

Quantum jumps as described above emerged from quantum trajectory theory
in quantum optics for open quantum systems where the environment is comprised of electromagnetic fields (a collection of harmonic oscillators) \cite{Car08, DalMol92, GarZol92, Bar93, GisPer92b, WisMil93}. In bosonic-environment settings, the environment is often treated as being in the vacuum state (analogous to a qubit ground state) or extremely close to vaccum in that each weak interaction time possesses only a small single-photon probability, $\wp \sim \mathcal{O}(\gamma dt)$~\cite{Baragiola:2012aa, Baragiola:2017aa, Gross:2021tv}. That is, the probability of a single photon in the field in each time step is proportional to $\Delta t$, which becomes infinitesimal as $\Delta t \rightarrow dt$. This is in marked contrast to a (stream of single) qubit environment model, where each qubit can be prepared in an arbitrary superposition or mixture of ground and excited states with no limit on the excited state amplitude. 

The other type of commonly considered conditional dynamics is diffusive evolution. Consider the projection of the bath qubit in our example onto the $X$-basis eigenstates, 
\begin{equation} \label{pmbasis}
    \ket{\pm} = \frac{1}{\sqrt{2}}(\ket{\er} \pm \ket{\gc}),
\end{equation}
with two outcomes $\pm$ that have nearly the same probability: $\wp_{\pm} = \smallfrac{1}{2} \pm \sqrt{\gamma} \expt{\hat c+ \hat{c}\dg} \sqrt{\Delta t} \sim \smallfrac{1}{2} \pm O(\sqrt{ \gamma \Delta t})$. The associated conditional maps are only perturbatively different from the identity map, such that the conditional state of the system is not significantly modified at each measurement time and neither outcome provides much system information to an observer~\cite{Gross:2021tv}. As these measurements are repeated, and in the continuous-time limit~\cite{Pellegrini:2009wz}, the system state undergoes diffusive dynamics, in sharp contrast to the jump-type evolution described above.

An interesting observation is that the nature of the conditional evolution (and thus the quantum trajectories) depends on how the measurement basis relates to the initial environmental state $\ket{\psi_B}$. Jump-type evolution can arise when $\ket{\psi_B}$ has no overlap with one of the outcomes, and diffusive evolution can arise when $\ket{\psi_B}$ has equal overlap with the outcomes. 

The two types of dynamics discussed above arise when the bath measurements are two outcome---\emph{i.e.} when the bath is a single qubit.
More generally, input environmental states and/or measurement bases can give rise to richer dynamics that mix jump- and diffusion-type dynamics, which are readily studied with repeated interaction models \cite{Pel10b}. 
In this work, we find that two-qubit baths can drive a new type of dynamics that we call \emph{quasi-diffusive} quantum trajectories. Quasi-diffusive dynamics are different from true diffusive dynamics, and they do not arise with more commonly studied two-mode bosonic baths in quantum optical settings. There, the white-noise conditions require that the joint probability of detecting more than one photon always vanishes, even when multiple bath modes are considered. For instance, over two bath modes, the joint two-photon detection probability scales as $\wp_\text{two-jump} \sim \mathcal{O}[(\gamma dt)^2] = 0$. A consequence is that jump-type dynamics always has a single no-jump map with probability near 1 (no photons detected) and possibly many jump-type maps, each corresponding to a single photon detected in a single mode.
In contrast, qubit baths do not have this restriction---the probability of detecting the bath qubits in $\ket{\er \er}$ can be non-negligible or even significant. As we will see, for two-qubit baths, this can split the no-jump dynamics into two separate maps, each with significant probability. Thus, most of the time (\eg in the absence of jumps), one of the two no-jump maps is applied, randomly chosen from the two, in analogy to true diffusive trajectories. However, the nontrivial terms in the Kraus operators scale as $\mathcal{O}(\gamma \Delta t)$---instead of $\mathcal{O}(\sqrt{\gamma \Delta t})$ for true diffusive maps---thus the effects are less pronounced.

\section{Master equations for entangled two-qubit baths} \label{sec:MEsection}

The focus of this work is the situation described in Fig.~\ref{fig:setup}, where two distinct, fixed, and potentially remote quantum systems repeatedly interact with identically prepared two-qubit environments, which are then detected. Within the formalism, the systems can be two-level systems (qubits) themselves, or they can be more exotic objects such as bosonic modes or atomic-type systems with complex internal structure. The dynamics of such a system coupled to two-qubit baths is quite rich---depending on the state of the bath qubits, the system can experience completely distinct evolution as analyzed in \crf{DarGil18} when the bath qubits are traced out. The reduced-system dynamics in this case are described by an unconditional master equation. We briefly review the results of that work including the specific entangled two-qubit bath states considered, as we will be building on those results throughout this work.

\subsection{State of the bath qubits}

Following \crf{DarGil18}, we consider a two-qubit state for the repeated environment $\hat{\rho}_{B}$ described by the block-diagonal density matrix in the basis $\{\ket{\er\er}, \ket{\gc\gc}, \ket{\er\gc}, \ket{\gc \er} \}$:
\beq  \label{rhoE:entgen}
\hat{\rho}_{\rm B} =  \left( 
\begin{array}{cccc}
\abs{b_{\er\er}}^2  & b_{\er\er} b_{\gc\gc}^*  & 0 & 0 \\ 
 b_{\er\er}^* b_{\gc\gc} & \abs{b_{\gc\gc}}^2  & 0 &  0\\
0 & 0 & \abs{b_{\er\gc}}^2 &  b_{\er\gc} b_{\gc\er}^*   \\
0 & 0 &b_{\er\gc}^* b_{\gc\er}  & \abs{b_{\gc\er}}^2  
\end{array}
\right), 
\eeq
with $ \tr{\hat{\rho}_{\rm B}}=1$.\footnote{This two-qubit state appears different from that in \crf{DarGil18}, because the basis vectors there are ordered differently.}
While not the most general state, this two-qubit state can exhibit quantum correlations between the two qubits, which is the essential feature in the quantum trajectories. 
This can be seen by recognizing that the state is a mixture of two orthgonal pure states, one in each of two orthogonal subspaces, $\text{span}\{\ket{\er\er},\ket{\gc\gc}\}$ and $\text{span}\{\ket{\er\gc},\ket{\gc\er} \}$:
\bqa \label{qubitstatedecomp}
	\hat{\rho}_B =  p_\typea \op{\psi_{B_\typea}}{\psi_{B_\typea}} +  p_\typeb \op{\psi_{B_\typeb}}{\psi_{B_\typeb}}
	\, ,
\eqa
with $\ip{\psi_{B_\typea}}{\psi_{B_\typeb}} = 0$. Drawing the coefficients from the density matrix above, the normalized states are
\bse \label{inputBellstates}
\bqa
\ket{\psi_{B_\typea}} &=&  \frac{1}{\sqrt{p_\typea}} \big( b_{\er\er} \ket{\er\er}+ b_{\gc\gc} \ket{\gc\gc} \big), \label{eq:state1} \\
\ket{\psi_{B_\typeb}} &=& \frac{1}{\sqrt{p_\typeb}} \big( b_{\er\gc}  \ket{\er\gc}+ b_{\gc\er} \ket{\gc\er} \big), \label{eq:state2}
\eqa  
\ese 
and the corresponding mixture probabilities for the Bell subspaces are
\bse \label{prob:mixed:rho}
\bqa
p_\typea &=&  \abs{b_{\er\er}}^2+\abs{b_{\gc\gc}}^2\, , \\
p_\typeb &=& \abs{b_{\er\gc}}^2+\abs{b_{\gc\er}}^2
\, ,
\eqa  
\ese
with $p_\typea + p_\typeb = 1$.\footnote{Take note that $b_{\gc \er}$ and $b_{\er \gc}$ are amplitudes in the pure state, \erf{eq:state2}, not coherences in a density matrix, so in general $b_{\gc \er} \neq b^*_{\er \gc}$.}

A useful observation is that each subspace is also spanned by two maximally Bell states, defined as
	\begin{subequations} \label{Bellstates}
	\begin{align}
		\ket{\Phi^\pm} &\coloneqq \smallfrac{1}{\sqrt{2}}\( \ket{\er\er} \pm \ket{\gc\gc} \), \label{Bell:phi:pm} \\
		\ket{\Psi^\pm} &\coloneqq \smallfrac{1}{\sqrt{2}}\( \ket{\er\gc} \pm \ket{\gc\er} \). \label{Bell:psi:pm}
	\end{align}
	\end{subequations}	
Thus, $\ket{\psi_{B_\typea}} \in \text{span}\{\ket{\Phi^+},\ket{\Phi^-}\}$ and $\ket{\psi_{B_\typeb}} \in \text{span}\{\ket{\Psi^+},\ket{\Psi^-}\}$.
By selecting the coefficients $b_{jk}$ appropriately, we can use the state in \erf{rhoE:entgen} to describe all four Bell states as well as various superpositions and mixtures thereof.

 \subsection{Master equation} \label{sec:ME}
In Ref.~\cite{DarGil18}, a Markovian master equation for the reduced system state $\hat\rho$ was obtained by evaluating the dynamical map, \erf{map:gen}, for the above two-qubit bath state in an appropriate continuous-time limit. We briefly review that result, as it will give insight into the quantum trajectories that are the main focus of this work. Additionally, the master equation and quantum trajectories are connected by the fact that an average over measurement outcomes given any measurement scheme (including the two discussed below), yields the same master equation. 

The Gorini–Kossakowski–Sudarshan–Lindblad master equation, commonly referred to simply as a Lindblad master equation, is~\cite{DarGil18}
\begin{align} \label{ME:gen}
\dot{\hat \rho}(t) 
& =  \sum_{m=1}^4 \mathcal{D}[ \hat L_m] \hat \rho \coloneqq {\cal L} \hat \rho ,
\end{align}
with Lindblad operators,\footnote{Often, these are also called ``jump operators,'' but we choose not to use that term, because of the frequent use of the word ``jump'' for other things.}
\begin{subequations}  \label{eq:jumpops}
	\begin{align}
	\hat L_1 =& \sqrt{ \g{} } \big( b_{\gc \gc} \hat{c}_1 + b_{\er \er} \hat{c}_2\dg  \big),  \\
	\hat L_2 =& \sqrt{ \g{} } \big( b_{\er \er} \hat{c}_1\dg +  b_{\gc \gc} \hat{c}_2 \big), \\
	\hat L_3 =& \sqrt{ \g{} } \big( b_{\gc \er} \hat{c}_1 +  b_{\er \gc} \hat{c}_2 \big), \\
	\hat L_4 =& \sqrt{ \g{} } \big( b_{\er \gc} \hat{c}_1\dg +  b_{\gc \er} \hat{c}_2\dg \big), 
	\end{align}
\end{subequations}
and the dissipation superoperator is
\beq \label{eq:Lindbladdissipator}
{\cal D}[\hat o] \hat{\rho} \coloneqq \hat o \hat \rho \hat o\dg - \frac{1}{2} \(\hat \rho \hat o\dg \hat o +  \hat o\dg \hat o \hat \rho\).
\eeq
We have set the rates, \erf{weakcouplingrate}, equal in the Lindblad operators, $\gamma_1 = \gamma_2 = \gamma$, and henceforth we make this simplification in all formulae in order to clarify expressions.\footnote{To restore separate rates in any of the equations, use the following simple procedure. First, remove all instances of the rate $\gamma$, then make the replacements $\hat{c}_\ell \rightarrow \sqrt{\gamma_\ell} \hat{c}_\ell$.} 

This master equation drives incoherent dynamics on the system via four decoherence channels each of which is responsible for a correlated dissipative process that is a combination of loss $(\hat c_\ell)$ and heating $(\hat c_\ell\dg)$ across both subsystems.  
Finally, it is interesting that the Lindblad operators $\hat L_1$ and $\hat L_2$ are determined by the portion of the environmental state in the Bell subspace $\{\ket{\er\er},\ket{\gc\gc}\}$ and the last two Lindblad operators, $\hat L_3$ and $\hat L_4$, by the portion of the state in the other Bell subspace $\{\ket{\er\gc},\ket{\gc\er}\}$. Further details can be found in Ref.~\cite{DarGil18}. These master-equation Lindblad operators will be useful for more compactly representing some of the Kraus operators and conditional maps in the remainder of this work.

We will find it useful later to rewrite the master equation, \erf{ME:gen}, by expanding each Lindblad dissipator, \erf{eq:Lindbladdissipator}, and regrouping terms to give, 
\beq  \label{fbME}
\dot{\hat\rho}(t) = -i 
\big( \hat H_\text{eff} \hat\rho - \hat\rho \hat {H}_\text{eff}\dg \big) + \sum_{m=1}^4 {\cal J}[\hat L_m] \hat\rho .
\eeq 
In this form, the dynamical map is divided into a portion generated by an effective anti-Hermitian Hamiltonian,
\beq \label{eq:nonHH}
\hat H_\text{eff} = -\frac{i}{2} \sum_{m=1}^4 \hat L_m\dg \hat L_m,
\eeq
and a portion described by the superoperator 
\begin{equation} \label{MEjumpop}
 {\cal J}[\hat o] \hat\rho \coloneqq \hat o \hat\rho \hat o\dg .
\end{equation}
This division is a generic feature of Lindblad-form master equations and can be useful for understanding the connections quantum trajectories, as we see in the following sections. 

We present the eigenvalues and eigenstates of $\hat H_\text{eff}$, which will be useful later. 
$\hat H_\text{eff}$ has spectrum
\begin{align} \label{eval:nonHH:genero}
{\mathbf w} 
& =
-i \tfrac{\gamma}{2} \left(
p_\typeb ,
p_\typeb + 2p_\typea ,
p_\typea ,
p_\typea + 2p_\typeb
\right), 
\end{align}
with $p_j$ defined in \erf{prob:mixed:rho}. The associated orthonormal eigenstates are
\bse \label{evec:nonHHgeneigvec}
\bqa  
\ket{\text{w}_1} &=& \tfrac{1}{\sqrt{p_\typea}} \big( b_{\er \er}
\ket{\er\er} - b_{\gc \gc} \ket{\gc \gc} \big),
\\
\ket{\text{w}_2} &=& \tfrac{1}{\sqrt{p_\typea}} \big( b^*_{\gc \gc}
\ket{\er\er} + b^*_{\er \er} \ket{\gc \gc} \big) ,
\\
\ket{\text{w}_3} &=& \tfrac{1}{\sqrt{p_\typeb}} \big( b_{\er \gc} \ket{\er \gc} - b_{\gc \er} \ket{\gc \er} \big),
\\
\ket{\text{w}_4} &=& \tfrac{1}{\sqrt{p_\typeb}} \big( b^*_{\gc \er}
\ket{\er\gc} + b^*_{\er \gc } \ket{\gc \er} \big), 
\eqa 
\ese 
such that the state $\hat \rho_{jk} = \op{{\rm w}_j}{{\rm w}_k}$ is an eigenoperator of evolution under the effective Hamiltonian alone: $-i\big( \hat H_\text{eff} \hat \rho_{jk} - \hat \rho_{jk} \hat {H}_\text{eff}\dg \big) = -i (\text{w}_j - \text{w}_k^\ast) \hat \rho_{jk}.$
Note that the two groups of eigenstates, $\{ \ket{ \text{w}_1}, \ket{ \text{w}_2} \}$ and $\{ \ket{ \text{w}_3}, \ket{ \text{w}_4} \}$, span orthogonal subspaces. 
\blk

A brief comment is necessary here for the case when both coefficients vanish in one of the two subspaces, such as $b_{\er \gc} = b_{\gc \er} = 0$ ($p_\typeb = 0$). In this case, two of the Lindblad operators vanish, $\hat{L}_3=\hat{L}_4 =0$, and the effective Hamiltonian is diagonal in the $\ket{\Psi^{\pm}}$ subspace.
Rather than using the equations above for $\ket{\text{w}_3}$ and $\ket{\text{w}_4}$, 
one should manually construct the remaining eigenvectors that span the trivial subspace (associated with degenerate eigenvalue $\text{w}_3 = \text{w}_4 = 1$) if a complete set is required. 
The spanning states can be chosen arbitrarily as long as they respect orthonormality. Two choices for $\{ \ket{\text{w}_3}, \ket{\text{w}_4} \}$ we make use of below are
$\{ \ket{\er \gc}, \ket{\gc \er} \}$ and $\{ \ket{\Psi^+}, \ket{\Psi^-} \}$.

\section{Quantum trajectories for entangled two-qubit baths} \label{sec:quantumtrajectoriesEB}

We extend the work of Ref.~\cite{DarGil18} to find the quantum trajectories for the conditional system state when the bath qubits are measured in various ways.  
For various measurement bases, we present the necessary components for quantum trajectories---most critically the weak-measurement Kraus operators, \erf{KrausWeak}, from which the conditional maps, outcome probabilities, and conditional difference equations can be derived.
We devote much of this section to local energy-basis measurements, where each bath qubit is measured individual in the $\{\er, \gc\}$ basis. 
Kraus operators (and associated quantum trajectories) for other measurement bases can be constructed using linear combinations of those for local energy-basis measurement. We give two examples: entangled Bell-basis measurements that gives jump-type dynamics and local basis measurements where one of the qubits is measured in $\ket{\pm}$, giving rise to diffusive dynamics. 

The master-equation analysis above considered a mixed bath state, \erf{inputBellstates}. For the quantum trajectories, we focus on the pure state
\begin{equation} \label{inputbathstate}
    \ket{\psi_{B_\typea}} =  b_{\er\er} \ket{\er\er}+ b_{\gc\gc} \ket{\gc\gc} ,
\end{equation}  
which is just \erf{eq:state1} with $p_{\Phi} = | b_{\er\er}|^2 + | b_{\gc\gc}|^2 = 1$.
We leave the amplitudes $b_{\er\er}$ and $b_{\gc\gc}$ unspecified, as their values can significantly influence the conditional dynamics. For the interested reader, in Appendix~\ref{appn_preA} we present Kraus operators, conditional maps, and probabilities for the other state, \erf{eq:state2}.

\subsection{Local energy-basis measurements}  \label{sec:maps:gen}
 
 There are infinitely many detection schemes that can be used for projective measurements of the environment qubits. Each detection scheme---characterized by the basis in which the environmental qubits are measured---leads to a different map and type of conditional system evolution (two examples are jump-type and diffusive-type as discussed in Sec.~\ref{ConditionalTypes}). This measurement-dependence is lifted if an ensemble average is taken over the outcomes; doing so produces the master equation in \erf{ME:gen} in the continuous-time limit regardless of the measurement basis.

When each of the bath qubits is measured in its local eigenbasis, the four measurement outcomes correspond to projections onto the following basis set:
\beq \label{MB:local}
{\bf m} = \cu{\ket{\er\er},\ket{\gc\gc}, \ket{\er\gc}, \ket{\gc\er} }
\, ,
\eeq
whose elements we label as $m_j$.
The first step towards finding conditional maps is to work out the corresponding Kraus operators.

\subsubsection{Kraus operators} \label{loc:basis:theory}

In jump-type quantum trajectories, the Kraus operators fall into two different categories: those where the input state and the outcome are in the same subspace (no-jump Kraus operators), and those where they are in different subspaces (jump Kraus operators). As we will see, these two types of Kraus operator lead to different types of conditional map.

Consider the outcomes $\{\er\er\}$ or $\{\gc\gc\}$, which are in the same subspace as the input state. These yield no-jump Kraus operators, \erf{KrausWeak}, which are found by respective projections onto $m_1 = \ket{\er\er}$ and $m_2 = \ket{\gc\gc}$,
\begin{subequations} \label{eq:Krauspsi1nojump}
\begin{align} \label{Kraus:v1:1}
\hat{K}_{m_1}\typeasuperscript
   & = b_{\er\er} \Big( \hat{I}_S - \frac{\g{} \Delta t}{2} \sum_{\ell=1,2} \hat{c}_\ell \hat{c}_\ell\dg \Big) - b_{\gc\gc} \g{} \Delta t \hat{c}_1 \hat{c}_2 \, ,\\
   \hat{K}_{m_2}\typeasuperscript 
   & = b_{\gc\gc} \Big( \hat{I}_S - \frac{\g{} \Delta t}{2} \sum_{\ell=1,2} \hat{c}^\dagger_\ell \hat{c}_\ell \Big) - b_{\er\er} \g{} \Delta t  \hat{c}^\dagger_1 \hat{c}^\dagger_2
   \, , \label{Kraus:v1:4}
\end{align}
\end{subequations}
with equal effective rates for each subsystem $\gamma = \lambda^2 \Delta t$, arising from $\lambda_1 = \lambda_2 = \lambda$ in \erf{weakcouplingrate}, and we have kept terms to order $\gamma \Delta t$ according to weak coupling.
The expressions for the outcome probabilities are lengthy and are presented in Appendix \ref{appnA:loc}. It is useful to inspect the scaling of the probabilities\footnote{Note that the scaled probabilities might not necessarily add up to one. This is because only those probability amplitudes that change the scaling are kept in the expressions.},
 	\begin{subequations} \label{eq:state1probscaling}
	\begin{align} 
		\wp_{m_1} & \sim \abs{b_{\er\er}}^2 - \abs{b_{\er\er}} \mathcal{O}(\gamma \Delta t) ,\\
		\wp_{m_2} & \sim \abs{b_{\gc\gc}}^2 - \abs{b_{\gc\gc}} \mathcal{O}( \gamma \Delta t),
	\end{align}	 
	\end{subequations}
as different bath amplitudes can drastically alter the nature of the probabilities and the conditional maps.

We take a moment to discuss the form of these Kraus operators and their associated probabilities, focusing on $\hat{K}_{m_1}\typeasuperscript$, with all arguments likewise applying $\hat{K}_{m_2}\typeasuperscript$ under complementary settings.
First, consider the portion multiplied by $b_{\er\er}$. This portion is dominated by the first term (proportional to identity), with the outcome-dependent terms being a perturbative correction at order $\g{} \Delta t$. This is the form of a standard ``no-jump'' Kraus operator. The other portion of \erf{Kraus:v1:1}, proportional to $b_{\gc\gc}$, arises from the $\s{1}^\dagger \s{2}^\dagger$ terms in $(\hat{H}_{SB})^2$. This term has no counterpart in quantum optical settings, because the analogous terms that simultaneously create or destroy two photons in the same infinitesimal interval vanish~\cite{Baragiola:2017aa}.\footnote{The statement is equivalent to the Ito table in QSDE treatments~\cite{Gardiner85}.} 
The balance between these portions can be tuned by adjusting the relative amplitudes in the bath-qubit state, $|b_{\er\er}|$ and $|b_{\gc\gc}|$. 

For even a small amount but nonvanishing $|b_{\er\er}|$, the identity term in $\hat{K}_{m_1}\typeasuperscript$ is the dominant one, and the conditional state undergoes a ``no-jump'' type evolution, where it is only perturbatively disturbed. We remark on an interesting difference from the quantum optical case, where due to the white-noise limit, only a single outcome, occurring with probability near 1, gives a no-jump type Kraus operator. Here, two separate outcomes, which can have significant and similar probabilities (most pronounced at $b_{\er \er} = b_{\gc \gc} = \frac{1}{\sqrt{2}}$), give no-jump Kraus operators. During a system's conditional evolution, most often one of these two randomly occurs---the jumps, described below, only occur with probabilities $\sim \gamma \Delta t$.
This is reminiscent of diffusive quantum trajectories, but it differs because the Kraus operators here no-jump type, \emph{not} the diffusive type, whose nontrivial terms are of order $\sqrt{\gamma \Delta t}$ compared to $\gamma \Delta t$ here. Since $\sqrt{\gamma \Delta t} \gg \gamma \Delta t$, true diffusive trajectories exhibit much more significant random behavior, which is why we call these random no-jump dynamics \emph{quasi-diffusive}.

An important threshold occurs when $b_{\er \er} \sim \sqrt{\gamma \Delta t}$.\footnote{One might at first think that $b_{\er \er} \sim \sqrt{\gamma \Delta t}$ yields an analogue to quantum optical models, as a similar situation describes single-mode setting with single-qubit baths~\cite{GroCom17}. For example, a continuous-wave coherent state of amplitude $\alpha$ (with $|\alpha|^2$ giving the photon flux) has a continuous tensor-product decomposition $\ket{\alpha} = \bigotimes_t \ket{\alpha_t}$. Each $\ket{\alpha_t}$ is an infinitesimal decomposition at time $t$~\cite{Baragiola:2017aa, Gross:2021tv}, where the state is expanded in a basis of vacuum and a single photon with higher order terms vanishing; here, the unnormalized state is $\ket{\alpha_t} = \ket{0_t} + \alpha \sqrt{ dt} \ket{1_t}$.
However, that is a description in a single mode only. Across two modes, the two-photon probability vanishes in the white-noise limit. Thus, the qubit description of that situation has $b_{\er \er} = 0$.}
Then, $b_{\gc \gc} \sim 1 - \frac{1}{2} \gamma \Delta t$, and the no-jump Kraus operators are
\bse 
\begin{align} \label{smalleeKraus}
\hat{K}_{m_1}\typeasuperscript
   &\rightarrow  \sqrt{\gamma \Delta t} \hat{I}_S ,
   \\
   \hat{K}_{m_2}\typeasuperscript 
   &\rightarrow \Big(1 - \frac{1}{2} \gamma \Delta t \Big) \hat{I}_S - \frac{\g{} \Delta t}{2} \sum_{\ell=1,2} \hat{c}^\dagger_\ell \hat{c}_\ell .
\end{align}
\ese 
The outcome $\{\er \er \}$ occurs with probability $\sim \gamma \Delta t$, and its effect on the system is trivial. 
If $b_{\er \er}$ is any smaller, \emph{i.e.} $b_{\er \er} \sim \gamma \Delta t$ or the limiting case $b_{\er \er} = 0$, something interesting happens in the conditional evolution. Applying $\hat{K}_{m_1}\typeasuperscript$ as in \erf{map:cond} generates a conditional state proportional to $(\gamma \Delta t)^2$, which is vanishing in the weak-coupling limit.
The remedy to the apparent problem is that the probability for this outcome, \erf{prob:cond}, also vanishes at the same order as is evident in \erf{eq:state1probscaling}. What this means is that the outcome $\{\er\er\}$ never occurs when the bath state is perturbatively close to $\ket{\gc \gc}$. 

The two outcomes in the subspace orthogonal to $\ket{\psi_{B_\typea}}$, $\{\er \gc \}$ and $\{\gc \er\}$, give jump dynamics. Projections onto $m_3 = \ket{\er\gc}$ and $m_4 = \ket{\gc\er}$ give the respective weak-measurement Kraus operators,
\begin{subequations} \label{eq:Krauspsi1jump}
	\bqa
	\hat K_{m_3}\typeasuperscript &=&-i \sqrt{ \g{}\Delta t}  \big(b_{\gc\gc}  \hat{c}_1 + b_{\er\er}  \hat{c}_2\dg \big) = -i \sqrt{ \Delta t} \, \hat{L}_1 , \\
	\hat K_{m_4}\typeasuperscript &=& -i \sqrt{\g{}\Delta t} \big(b_{\er\er} \hat{c}_1\dg + b_{\gc\gc} \hat{c}_2 \big) = -i \sqrt{ \Delta t}\, \hat{L}_2  .
	\eqa
\end{subequations}
On the right-hand side of each equation, we have expressed the Kraus operators in terms of Lindblad operators $\hat{L}_i$ that arise in the master equation, \erf{eq:jumpops}. 
We retain the $-i$ phases, because superpositions of these Kraus operators can be used to describe other measurement bases where such phases can matter. These Kraus operators have no component proportional to the identity, so they give rise to ``jump-type'' conditional maps that significantly change the system state. Since each is the coherent sum of two operators, they can generate superpositions or, if the system itself is comprised of two subsystems, entanglement. The probability of the outcomes to occur, \erf{prob:cond}, scale as
	\begin{equation} \label{eq:state1probscaling34}
	\wp_{m_3}, \wp_{m_4} \sim \mathcal{O}(\gamma \Delta t),
	\end{equation}
making them much more infrequent than the ``no-jump'' outcomes. Unlike \erf{eq:state1probscaling}, there is no dependence on $b_{\er \er}$ and $b_{\gc \gc}$ in the order of the scaling---it's $\mathcal{O}(\gamma \Delta t)$ regardless.

We have presented the full set of four Kraus operators, with each corresponding to an outcome in the local energy basis across both bath qubits. Together they resolve the identity,
	\begin{align} \label{eq:resolveidentity}
		\sum_{j=1}^4 \hat K_{m_j}\typeasuperscript^\dagger \hat K_{m_j}\typeasuperscript 
= \hat{I}_S
,
	\end{align}
thus, the Kraus operators comprise a valid POVM, and the sum over outcome probabilities is equal to 1. 

\subsubsection{Conditional maps}

The Kraus operators are the fundamental objects used to describe quantum trajectories by simply updating the state according to the conditional map, \erf{map:cond}, once an outcome is obtained. 
These equations are lengthy and can be found in Appendix~\ref{appnA:loc} along with expressions for the outcome probabilities.
If the bath is in the mixed state described by \erf{qubitstatedecomp} with $0< p_\typea, p_\typeb < 1$, then the appropriate map for a given measurement outcome $m_j$ for any measurement basis is the convex sum of the maps for each pure state [Eq.~\eqref{map:condgenmix}],
\beq \label{map:mixed}
\vo(t_{k +1}) 
= \frac{1}{\wp_{m_j}} \big[ p_\typea \mathcal{E}_{m_j}\typeasuperscript ( \hat \varrho ) + p_\typeb \mathcal{E}_{m_j}^{(\typeb)} ( \hat \varrho )  \big]
\, ,
\eeq	
and the probability of obtaining the outcomes is 
\begin{equation} \label{cond:rho:gen}
	 \wp_{j} = p_\typea \text{Tr} \big[ \hat{\varrho} \hat K_{m_j}^{(\typea)\dagger} \hat K_{m_j}\typeasuperscript \big] + p_\typeb \text{Tr} \big[ \hat{\varrho} \hat K_{m_j}^{(\typeb)\dagger} \hat K_{m_j}^{(\typeb)} \big]
	 \, .
\end{equation}
	
Finally, averaging the conditional maps using the outcome probabilities supplies the connection to the master equation, \erf{ME:gen}, 
\begin{align} \label{eq:mapconnect}
\sum_{j=1}^4 \Big[ p_\typea \hat{K}_{m_j}\typeasuperscript \hat \varrho \hat{K}^{(\typea)\dagger}_{m_j} + p_\typeb \hat{K}_{m_j}^{(\typeb)} \hat \varrho \hat{K}^{(\typeb)\dagger}_{m_j} \Big]
- \hat \varrho = \Delta t \mathcal{L} \hat \varrho
.
\end{align}

\subsubsection{Conditional difference equations}

With the conditional maps, one can also construct conditional difference equations, \erf{dffmap}, which can be instructive for analysis and for comparison with other types of quantum trajectory equations. 
The general forms for the conditional difference equations are straightforward using the conditional maps and probabilities, which are given in detail in Appendix~\ref{appnA:loc}. We present a useful form for the conditional difference equations that is valid when the probability of getting both no-jump outcomes is significantly greater than $\gamma \Delta t$---that is, when $|b_{\er \er}|, |b_{\gc \gc}| > \sqrt{\gamma \Delta t}$. In this case, the jump equations take the general form, but the no-jump equations can be re-expressed in a form typical of standard SMEs. Their derivation is given in Appendix~\ref{appn_conddiffeqs}, and the resulting set of conditional difference equations is,
\begin{subequations} \label{genconddiffs}
\begin{align} 
\Delta \vo_{m_1} 
&=  
-\g{} \Delta t \Big( \sum_{j=1,2} \tfrac{1}{2} {\cal M} \big[ \hat{c}_j \hat{c}^\dagger_j \big] +
  {\cal M} \Big[ \frac{b_{\gc \gc}}{b_{\er \er}} \hat{c}_1 \hat{c}_2 \Big]  \Big)\vo,  \label{drho:loc1:nmax}  \\
\Delta \vo_{m_2} &= -\g{} \Delta t \Big(  \sum_{j=1,2} \tfrac{1}{2}  {\cal M} \big[ \hat{c}^\dagger_j \hat{c}_j \big] + {\cal M}  \Big[ \frac{b_{\er \er}}{b_{\gc \gc}} \hat{c}^\dagger_1 \hat{c}^\dagger_2 \Big] \Big)\vo,  \label{drho:loc2:nmax} \\
 \Delta \vo_{m_3} &=  \mathcal{G}[ b_{\gc \gc} \hat{c}_1 + b_{\er \er} \hat{c}_2\dg ] \hat \rho = \mathcal{G}[\hat{L}_1] \hat \rho ,  \label{drho:loc3:nmax}\\
 \Delta \vo_{m_4} &=  \mathcal{G}[ b_{\er \er} \hat{c}_1\dg + b_{\gc \gc} \hat{c}_2] \hat \rho = \mathcal{G}[\hat{L}_2] \hat \rho  , \label{drho:loc4:nmax}
\end{align}
\end{subequations}
where $\Re[z]$ indicates the real part of complex number $z$, $\expt{ \hat o}= \text{Tr} [\hat o \hat{\varrho} ]$, and the standard no-jump and jump measurement superoperators are respectively defined
\begin{align} \label{eq:measurementsuperoperator}
{\cal M}[\hat o]\vo &\coloneqq \hat o \vo + \vo \hat{o}\dg - \expt{\hat o + \hat{o}\dg} \vo ,
\\
\mathcal{G}[\hat o] \hat \varrho &\coloneqq \frac{\hat{o} \hat \varrho \hat{o}^\dagger}{ \expt{\hat{o}^\dagger \hat{o}} } - \hat{\varrho} , \label{jump:superoperator}
\end{align}
The jump superoperator is related to the superoperator in the master equation, \erf{MEjumpop}: $\mathcal{G}[\hat o]\vo = \mathcal{J} (\hat o) \hat\varrho /\expt{\mathcal{J} [\hat{o}]} - \hat{\varrho}$.\footnote{It may seem curious that the two jump-type equations can be written in terms of the master-equation Lindblad operators $\hat{L}_1$ and $\hat{L}_2$, while the no-jump equations cannot. Although the no-jump Kraus operators do not individually factorize, they do \emph{average} to produce the anti-Hermitian dynamics in the alternate form of the master equation, 
\erf{fbME},
$\sum_{j=1,2}  \hat{K}_{m_j}\typeasuperscript \hat \varrho \hat{K}\typeasuperscriptdag_{m_j} - \hat \varrho = - i \Delta t \big( \hat{H}_\text{eff} \hat{\varrho} - \hat{\varrho} \hat{H}_\text{eff}^\dagger \big)
$. This becomes important in Sec.~\ref{3result:sme}.
Whether the no-jump maps can be written in terms of $\hat{L}_1$ and $\hat{L}_2$ depends on the measurement basis. In Sec.~\ref{Bellbasis_example}, we give an example using Bell-basis measurements where they can be.
} 
Following from the above analysis of the Kraus operators, if either $|b_{\er \er}|$ or $|b_{\gc \gc}|$ is too small, $\mathcal{O}(\sqrt{\gamma \Delta t}$), then the two no-jump equations, \erf{drho:loc1:nmax} and \erf{drho:loc2:nmax}, are not simultaneously valid, although one of them always is since $|b_{\gc \gc}|^2 + |b_{\er \er}|^2 = 1$. For example, when $|b_{\er \er}|$ is small, \erf{drho:loc2:nmax} is valid, but $\Delta \vo_{m_1} = 0$ is trivial [see \erf{smalleeKraus}]. In the limiting case, $|b_{\er \er}| = 0$, the outcome $\{\er \er \}$ never occurs at all, and the dynamics have only a single no-jump equation, just as in quantum optical settings.

\subsection{Other measurement bases: examples}

The main focus of this work are the quantum trajectories driven by local energy basis measurements analyzed above, since those Kraus operators can be used to generate any other set given by projective measurements of the bath qubits. Below, we highlight two other interesting measurement bases and provide some details of their quantum trajectories.

\subsubsection{Entangled Bell-basis measurements} \label{Bell:basis:theory}
As the natural couterpart to the local measurement basis, we also consider the situation where the two bath qubits are jointly measured in a maximally entangled basis correspoinding to projections onto the set
\begin{align} \label{MB:Bell}
    \{ \ket{\Phi^+}, & \ket{\Phi^-}, \ket{\Psi^+}, \ket{\Psi^-} \} \nonumber \\
    & = \frac{1}{\sqrt{2}} \cu{m_1 + m_2, m_1 - m_2, m_3 + m_4, m_3 - m_4} ,
\end{align}
where the Bell states are defined in \erf{Bellstates}. The second line gives the relation to the local energy basis in \erf{MB:local}, which allows us generate the Bell-basis Kraus operators with superpositions of the local-basis ones,
\bse 
\begin{align}
    \hat K_{\Phi^\pm} 
    = 
    \tfrac{1}{\sqrt{2}} \big(\hat K_{m_1}\typeasuperscript \pm \hat K_{m_2}\typeasuperscript\big) ,
    \\
    \hat K_{\Psi^\pm} 
    = 
    \tfrac{1}{\sqrt{2}} \big(\hat K_{m_3}\typeasuperscript \pm \hat K_{m_4}\typeasuperscript \big) ,
\end{align} 
\ese 
with explicit forms given in Appendix~\ref{appnA:joint}. 
The outcome probabilities scale as,
\bse 
    \begin{align}
        \wp_{\Phi^\pm} &\sim \smallfrac{1}{2}\abs{b_{\er\er}\pm b_{\gc\gc}}^2 - \abs{b_{\er\er}\pm b_{\gc\gc}} {\cal O}(\gamma \Delta t)
        \\
        \wp_{\Psi^\pm} &\sim \abs{b_{\er\er}\pm b_{\gc\gc}} {\cal O}(\gamma \Delta t).
    \end{align}
    \ese 
The superpositions in the Kraus operators still respect the division into the two Bell subspaces. That is, outcomes $\Phi^{\pm}$ are related only to $m_1$ and $m_2$; likewise $\Psi^{\pm}$ are related only to $m_3$ and $m_4$. Thus, the nature of the maps for each outcome, either jump or no-jump, are preserved. Similarly, the scaling of the probabilities and the forms of the conditional difference equations are preserved.

Although we do not delve further in this direction, one could consider more general cases, by measuring in an entangled basis of unequally weighted local states. In this case, the new Kraus operators would simply be appropriately weighted superpositions of the local-basis ones.

\subsubsection{Local XZ-basis measurements: towards diffusive dynamics} \label{sec:mixedlocalmeasurements}

Diffusive dynamics occur when a measurement basis is chosen such that the Kraus operators are superpositions of jump-type and no-jump type. This can be done in many different ways. For example, a diffusive Kraus operator could be constructed using a Kraus operator from \erf{eq:Krauspsi1nojump} with one from \erf{eq:Krauspsi1jump}. Following with other similar selections, the full set can be constructed such the POVM resolves the identity, \erf{eq:resolveidentity}.	
We take a measurement-model approach by making adjustments to local-basis measurements, as this is most relevant when the bath qubits themselves are spatially separated. 

We measure the first bath qubit in the $X$-basis, with outcomes $\{ \pm \}$ corresponding to projections onto $ \ket{\pm} $ in [\erf{pmbasis}], while the second qubit is still measured in the local energy basis $\{\er, \gc \}$, as above. We refer to this as \emph{local $XZ$-basis measurements}.
The measurement basis is spanned by the four bath-qubit states,
\begin{align} \label{MB:localdiff}
 \{ \ket{+ \, \er},& \ket{- \, \er}, \ket{+ \, \gc}, \ket{- \, \gc} \}  \nonumber \\  
 & = \frac{1}{\sqrt{2}} \cu{m_1 + m_4, m_1 - m_4, m_2 + m_3, -m_2 + m_3} ,
\end{align}
and the Kraus operators are found by combining the local energy-basis Kraus operators in \srf{loc:basis:theory} appropriately. 
For input state $\ket{\psi_{B_\typea}}$, \erf{inputbathstate}, 
the four Kraus operators are
\bse 
   \begin{align} \label{mixedlocalkraus}
        \hat{K}_{\pm\er}\typeasuperscript &= \tfrac{1}{\sqrt{2}} \big( \hat{K}_{m_1}\typeasuperscript  \pm \hat{K}_{m_4}\typeasuperscript\big) , \\
        \hat{K}_{\pm\gc}\typeasuperscript &= \tfrac{1}{\sqrt{2}} \big( \hat{K}_{m_3}\typeasuperscript  \pm \hat{K}_{m_2}\typeasuperscript\big) .
    \end{align}
    \ese 
with outcome probabilities scaling like,
\bse 
    \begin{align}
        \wp_{\pm\er} &\sim \frac{1}{2} \Big[ |b_{\er \er}|^2 \pm |b_{\er \er}| \mathcal{O}(\sqrt{\gamma \Delta t}) - \mathcal{O}(\gamma \Delta t)\Big]
        \\
        \wp_{\pm\gc} &\sim \frac{1}{2} \Big[ |b_{\gc \gc}|^2 \pm |b_{\gc \gc}| \mathcal{O}(\sqrt{\gamma \Delta t}) + \mathcal{O}(\gamma \Delta t) \Big].
    \end{align}
    \ese 
The explicit forms of the Kraus operators, the conditional maps they generate, and the probabilities are given in Appendix~\ref{appn:xzBasis}. The fact that there are four outcomes complicates a simple categorization of the dynamics just as it did for the local energy basis measurements above. That is because, while the first qubit's outcome drives diffusive-type dynamics, the second qubit's outcome sets this diffusion on one of two branches---that for $\{\er\}$ or that for $\{\gc\}$---which switch randomly. However, we show in Sec.~\ref{sec:diffusiveSME} that averaging over the second qubit's outcome yields a bonafide diffusive SME.

\section{Constructing stochastic master equations} \label{sec:eng:base}

In this section, we show that standard-form stochastic master equations (SMEs), such as those that arise in quantum-optical settings, can be derived from the above quantum trajectories. The recipes we give involve averaging over some of the four outcomes in the measurement models described above. However, as described below, the procedures by which a jump SME and a diffusive SME are derived do not involve the same type of average over outcomes. In the former, an average is performed over one outcome to yield a three-outcome measurement, and in the latter two averages are performed to yield an effective two-outcome measurement.

\subsection{Jump SME by averaging local energy-basis measurements} \label{3result:sme}

Consider a system driven by the two-qubit state $\ket{\psi_{B_{\typea} } }$, \erf{inputbathstate}. The system's unconditional, average dynamics are described by a master equation with Lindblad operators $\hat{L}_1$ and $\hat{L}_2$ from \erf{eq:jumpops}. In the alternate form, \erf{fbME}, the dynamical map in the master equation is comprised of two terms, ${\cal J}[\hat L_1] + {\cal J}[\hat L_1]$, and a commutator-type term containing the effective anti-Hermitian Hamiltonian, $\hat H_\text{eff} = -\frac{i}{2} ( \hat L_1\dg \hat L_1 + \hat L_2\dg \hat L_2 )$.
This suggests that there exists a three-outcome POVM with two jump Kraus operators 
\bse \label{3outcome-kraus:jump}
\begin{align}
\hat K_1 &= -i \sqrt{\Delta t} \,\hat L_1, \\
\hat K_2 &= -i \sqrt{\Delta t} \,\hat L_2,
\end{align}
\ese
and a no-jump Kraus operator
\begin{equation} \label{3outcome-kraus:nj}
\hat K_0 = \hat{I}_S - \Delta t (\smallfrac{1}{2}\hat L_1\dg \hat L_1+\smallfrac{1}{2}\hat L_2\dg \hat L_2) ,
\end{equation}
whose POVM elements satisfy $
\hat{I}_S = \sum_{j } \hat{K}_j\dg \hat{K}_j$. 

This three-outcome measurement setting can be realized by measuring in the local energy basis, \erf{MB:local}. 
One can immediately see that the two jump Kraus operators are those in \erf{eq:Krauspsi1jump}, $\hat K_1 = \hat K_{m_3}$ and $\hat K_2 = \hat K_{m_4}$. The third Kraus operator is found by discarding some information---specifically, which no-jump outcome occurred, $\{ \er \er\}$ or $\{ \gc \gc \}$. Then, the no-jump dynamics is given by averaging the two Kraus maps for $m_1$ and $m_2$ [using \erf{eq:Krauspsi1nojump}] according to their probabilities:
\beq \label{eq:mapsum}
\sum_{j=1,2} \wp_{m_j} \mathcal{E}_{m_j}(\vo)
= \sum_{j=1,2}  \hat K_{m_j} \vo \hat K_{m_j}^{\dagger} = \hat K_{0} \vo\, \hat K_{0}\dg 
.
\eeq
Typically, when averaging over outcomes in a measurement, the resulting map cannot be written as a single term in a Kraus representation. The reason that it can be here is that each of the individual Kraus maps separates into terms with operators acting from the left and terms with operators acting from the right. All cross-terms that would act from both sides in a sandwich fashion are higher order than $\mathcal{O}(\gamma \Delta t)$ and summarily vanish.
The measurement outcome probabilities are
\bse  \label{prob:3outcome:sme}
\begin{align}
    \wp_1 & 
    = \Delta t \expt{\hat L_1\dg \hat L_1} , \\
    \wp_2 & 
    =  \Delta t \expt{\hat L_2\dg \hat L_2}, \\
    \wp_0 & 
    =  1-(\wp_{1}+\wp_{2}), 
\end{align}
\ese
where the explicit forms can be found from the expressions in \erf{prob:loc:gen}. 
Since quantum jumps occur with probability in the order of $\mathcal{O}( \gamma \Delta t)$, we can describe their stochastic nature with two random variables described by Poisson increment $\Delta N_j$ for $j\in \{1,2\}$. Each $\Delta N_j$ can take on one of two values $\{0,1\}$, and satisfies
\bse 
\begin{align}
\left(\Delta N_j\right)^2 &= \Delta N_j \\
\label{mean:dN:3outcome}
{\mathbb E}\sq{\Delta N_j} &= \wp_j.
\end{align}
\ese 
Additionally, $\Delta N_1 \Delta N_2 = 0$, since the probability of both taking on the value 1 is $\mathcal{O}[(\gamma \Delta t)^2]$. These conditions are a reflection of that fact that the measurement model has three mutually exclusive outcomes. 

No-jump evolution occurs when both stochastic increments are zero, $\Delta N_1=\Delta N_2=0$, and a jump associated with Lindblad operator $\hat L_j$ occurs if $\Delta N_j=1$. With these random variables, we follow the strandard procedure and combine the conditional difference equations into a single SME:
\bse \label{diff:3outcome}
\bqa
\Delta \vo &=& \sum_{j=1,2}\Delta N_j \Delta \vo_{j}  + \Big(1 - \sum_{j=1,2}\Delta N_j\Big) \Delta \vo_{0},\\
&=& \sum_{j=1,2}\Delta N_j \Delta \vo_{j}  + \Delta \vo_{0},
\eqa
\ese
where the no-jump conditional difference equation is
\beq
\Delta \vo_{0} = \frac{\hat{K}_{0}\typeasuperscript \hat \varrho \hat{K}^{\dagger}_{0}}{\text{Tr} 
\big[ \hat{K}_{0}\typeasuperscript \hat \varrho \hat{K}^{\dagger}_{0} \big] } - \vo = -\Delta t \sum_{\ell=1,2} \smallfrac{1}{2} {\cal M}\big[\hat L_\ell\dg \hat L_\ell\big]\vo, \label{eq:jumpSME3outcome}
\eeq 
and for the jumps $\Delta \vo_j$ with $j=1,2$ are given in \erfa{drho:loc3:nmax}{drho:loc4:nmax}. 
In the second line of \erf{diff:3outcome}, terms containing $\Delta N \gamma \Delta t$ have been  discarded, because they are of vanishing order under expectation. This enforces that the jump and no jump outcomes do not simultaneously occur, which is simply a reflection of the fact that ours is a measurement model with three mutually exclusive outcomes. 

Equation (\ref{eq:jumpSME3outcome}) is a jump-type SME in standard form. An alternate form arises by using the \emph{jump innovations} $\Delta I_j \coloneqq \Delta N_j - {\mathbb E}\sq{\Delta N_j}$, which is a description of the new information gained in each measurement (expressed as a deviation from the expected value).
With this, the jump SME in \erf{eq:jumpSME3outcome} can be recast as
\begin{align} \label{diff:3outcome:Inn}
\Delta \vo 
&= \sum_{j=1,2} \Delta I_j\,{\cal G} \big[\hat L_j\big]\vo + \Delta t\, {\cal L}\vo
,
\end{align}
where ${\cal L}$ and ${\cal G}$ are defined in \erfa{ME:gen}{jump:superoperator} with $b_{\er\gc} = b_{\gc\er} = 0$ in the former. 
Taking a continuous-time limit in the appropriate way as discussed above in Sec.~\ref{sec:quantumtrajectories}, $\Delta t \rightarrow dt$, and this expression becomes the continuous-in-time SME:
\beq 
d \vo = \sum_{j=1,2} d I_j\,{\cal G}\big[\hat L_j\big]\vo + d t\, {\cal L}\vo
.
\eeq 
Here, $dI_j \coloneqq  dN_j - {\mathbb E}\sq{dN_j}$, where $dN$ represents a point process increment \cite{WisMil10}, whose expectation is given by \erf{mean:dN:3outcome} with $\Delta t$ replaced by $dt$.

As a final note, recall that the Bell-basis measurements in Sec.~\ref{Bell:basis:theory} respect the same division into jump and no-jump Kraus operators as the ones we consider here. Thus, the Bell-measurement maps also satisfy \erf{eq:mapsum}, with the sum running over $\Phi^\pm$. As a consequence, one can follow the same procedure by averaging over these two outcomes to produce an SME of the same form as \erf{diff:3outcome:Inn} where the jump portion $\mathcal{G}[ \hat{L}_\pm ]$ contains superpositions of the Lindblad operators, 
\begin{equation} \label{eq:Lpm}
   \hat{L}_\pm \coloneqq \tfrac{1}{\sqrt{2}} (\hat{L}_1 \pm \hat{L}_2).
   \end{equation}

\subsection{Diffusive SME by averaging local $XZ$ measurements} \label{sec:diffusiveSME}

Using a procedure similar to that above, we construct a diffusive SME from the local $XZ$-basis measurements in Sec~\ref{sec:mixedlocalmeasurements} by averaging over the second bath qubit's outcome. Unlike the previous section, this produces a two-outcome measurement with outcomes $\{ +, -\}$, since the outcomes on the second qubit are completely averaged out.

Using the Kraus operators, \erf{mixedlocalkraus}, we construct the average map for each outcome of the first qubit (+ or $-$),
\begin{align}
    &\mathcal{E}_\pm (\vo)
    \coloneqq \sum_{s\in\{\er,\gc\}} \hat{K}_{\pm s}\typeasuperscript\vo \hat{K}_{\pm s}\typeasuperscriptdag \nn
     \\
    &= \frac{1}{2} \Big\{\vo + \Delta t \sum_{j=1,2} {\cal D}\big[\hat L_j\big] \vo 
    \pm \sqrt{ \Delta t} ( \Ldiff \vo + \vo \Ldiffdg ) \Big\},
\end{align}
where we have defined
\beq 
\Ldiff \coloneqq -i\big(b_{\er\er}^* \hat L_1 + b_{\gc \gc}^* \hat L_2 \big).
\eeq 
The outcome probabilities $\wp_{\pm}$ are found by taking the trace of these expressions,
\begin{align}
    \wp_{\pm} 
    &= \frac{1}{2} \big( 1 \pm \sqrt{ \Delta t} \expt{\Ldiff + \Ldiffdg} \big)
    .
\end{align}
In contrast to the outcome averaging performed for the jump SME above, the averaging here does not result in a map described by a single Kraus operator. Thus, the resulting SME will have additional decoherence.
We now have two-outcome quantum trajectories, each outcome occurs with probability $\sim \frac{1}{2}$, and the associated maps contain outcome-dependent terms proportional to $\sqrt{\Delta t}$. These are the conditions for a standard diffusive-type SME.

Therefore, it is straightforward to calculate the following difference equations:
\bse \label{diff:eqs:XZ}
\begin{align} 
&\Delta \vo_\pm 
= \frac{\vo + \Delta t \sum_{j=1,2} {\cal D} \big[\hat L_j \big] \vo \pm \sqrt{\Delta t} (\Ldiff \vo + \vo \Ldiffdg  )}{1\pm \sqrt{\Delta t}  \expt{ \Ldiff + \Ldiffdg } } -\vo, 
\\
&\approx  \Delta t \sum_{j=1,2} {\cal D}\big[\hat L_j\big] \vo \pm \Big(  \sqrt{\Delta t} \mp \Delta t \expt{ \Ldiff + \Ldiffdg } \Big) {\cal M}\big[ \Ldiff \big] \vo,
\end{align}
\ese
where the superoperator $\mathcal{M}$ is defined in \erf{eq:measurementsuperoperator}. The second line follows by expanding the denominator and keeping resulting terms to $\mathcal{O}(\gamma \Delta t)$, similar to the procedure described in Appendix~\ref{appendix:conddiffenergybasis} for the no-jump conditional difference equations.

We now combine the two conditional difference equations into a diffusive-type SME. First, define a random variable $\Delta J$ that takes on values $\pm \sqrt{\Delta t}$ depending on the measurement result and has expectation 
\beq
{\mathbb E}\sq{\Delta J} = \sqrt{\Delta t}\, \wp_+ - \sqrt{\Delta t}\, \wp_- = \sqrt{\Delta t} \, \expt{ \Ldiff + \Ldiffdg }.
\eeq 
Introducing the \emph{diffusive innovations}
\beq 
\Delta W \coloneqq \Delta J - {\mathbb E}\sq{\Delta J}
,
\eeq 
with zero mean by definition and second moment
\beq
{\mathbb E} [ \Delta W ^2 ] 
= {\mathbb E}[ \Delta J^2 ] - \big({\mathbb E}\sq{\Delta J}\big)^2 
= \Delta t + {\cal O}\big[ (\gamma \Delta t)^2 \big].
\eeq
Thus, in the weak-coupling limit, $\Delta W$ is a Wiener increment.
With the diffusive innovations, one can group the two conditional difference equations in \erf{diff:eqs:XZ} into a single diffusive SME:
\beq  \label{SDE:XZ}
\Delta \vo = \Delta t \sum_{j=1,2} {\cal D}\big[\hat L_j\big] \vo + \Delta W \, {\cal M}\big[ \Ldiff \big] \vo . 
\eeq 
Note that this diffusive SME differs from single-channel versions, where the Lindblad operator that appears in the measurement superoperator ${\cal M}$ is the same that appears in the dissipator ${\cal D}$. This is because the averaging procedure we used to reduce the four-outcome quantum trajectories to two-outcome ones introduced additional decoherence.

The continuous-in-time form of \erf{SDE:XZ} can be constructed by taking the appropriate infinitesimal limit, which gives $\Delta t \rightarrow 0$ to obtain the following stochastic differential equation:
\beq 
d\vo = dt \, \sum_{j=1,2} {\cal D} \big[\hat L_j \big] \vo + dW\, {\cal M}\big[ \Ldiff \big] \vo,
\eeq 
where the Wiener process $dW$ is the continuous limit of the diffusive innovations satisfying $    {\mathbb E}[dW] = 0$ and ${\mathbb E} [dW^2] = dt$.
 
 \blk


\section{Example: Two remote two-level atoms interacting with  entangled two-qubit
baths} \label{2qubitsection}

The above quantum trajectories can be applied to arbitrary systems probed by entangled baths. 
We illustrate some properties of entangled-qubit-bath quantum trajectories by focusing on the specific situation where the bath couples to two remote subsystems, each a two-level atom. This situation was considered in Ref.~\cite{DarGil18} in the context of unconditional, master-equation dynamics.
Following that work---for the sake of clarity in discussions---we refer to the bath as \emph{qubits} and the subsystems as \emph{atoms} (which are themselves also qubits). After introducing the necessary details of the model, we review some features found in the master-equation analysis. Then, we lay out and analyz the quantum trajectories using different measurement bases: the local energy basis and the maximally entangled Bell basis.

We first describe the physical model. The two atom subsystems are identical with bare Hamiltonian $\hat{H}_{\ell} = \smallfrac{1}{2}\omega_{\er\gc} \,\hat{\sigma}_{z,\ell}$, where $\omega_{\er\gc}$ is the atomic transition frequency, and $\hat{\sigma}_z \coloneqq \op{\er}{\er} - \op{\gc}{\gc}$. Each atom interacts with its qubit bath through excitation exchange processes described by the atomic lowering and raising operators, $\hat{c}_\ell \rightarrow \s{\ell}$ and $\hat{c}^\dagger_\ell \rightarrow \s{\ell}^\dagger$. Just as for the general settings above, the couplings are equal $\lambda_\ell = \lambda$ in the interaction-picture Hamiltonian, \erf{IPIH:2q}, and the detuning is zero, $\delta_\ell=0$, indicating that the atoms and the bath qubits are energetically identical.

We focus on a specific set of pure bath qubit states that will allow us to write down conditional difference equations and explore situations that were introduced and analyzed in the case of unconditional master equations~\cite{DarGil18}.
Here, we consider bath states $\ket{\psi_B}$ described by two-qubit entangled states $\ket{\Phi_B^{+}(\epsilon)}$ in the subspace spanned by $\ket{\er\er}$ and $\ket{\gc\gc}\}$:
\begin{align} \label{phi:pm:ep}
\ket{\Phi^{\pm}(\epsilon)} & = \frac{1}{\sqrt{2+\epsilon}} \Big(\ket{\er\er} \pm \sqrt{1+\epsilon}\, \ket{\gc\gc} \Big).
\end{align}
This equation also defines $\ket{\Phi^{-}(\epsilon)}$, which will be useful below.
We call these \emph{near-Bell states}, because for $0 \leq \epsilon\ll 1$, they have fidelity $1 - \mathcal{O}(\epsilon^2)$ with their respective Bell state $\ket{\Phi^{\pm}}$ [\erf{Bell:phi:pm}], and they are exactly $\ket{\Phi^{\pm}}$ for $\epsilon = 0$. The near-Bell states are given by $p_\typea = 1$ and $p_\typeb = 0$ in \erf{inputBellstates}, with the amplitudes $b_{\er \er}$ and $b_{\gc \gc}$ being strictly real and parameterized by $\epsilon$.

Before we consider the various quantum trajectories, we briefly review some results for unconditional, master-equation dynamics under the bath qubit state $\ket{\Phi_B^{+}(\epsilon)}$. This can be useful in understanding the quantum trajectories, since the dynamics they describe average to the master-equation dynamics, independent of measurement basis. 
The average (unconditional) evolution of the two-atom state $\hat{\rho}$ is governed by the master equation
\begin{align}  \label{ME:Bell:phi}
	\dot{\hat{\rho}}(t) &=   \mathcal{D} \bigl[\hat L_1\bigr] \hat{\rho} +  \mathcal{D} \bigl[\hat L_2 \bigr] \hat{\rho}, 
\end{align}
with two Lindblad operators [see \erf{eq:jumpops}]:
\begin{subequations} \label{Lindblad:Bell:phi}
\begin{align} 
	\hat L_1 &= \sqrt{ \frac{\gamma}{2+\epsilon} } \big( \sqrt{1+\epsilon} \, \hat{\sigma}_1 + \hat{\sigma}^\dagger_2 \big ),
	\\
	\hat L_2 &= \sqrt{ \frac{\gamma}{2+\epsilon} } \big( \hat{\sigma}_1^\dagger + \sqrt{1+\epsilon} \, \hat{\sigma}_2 \big )
	.
\end{align}
\end{subequations}
In its alternate form, the master equation is
\begin{align}  \label{MEalt:Bell:phi}
 	&\dot{\hat{\rho}} = -i \big( \hat H_\text{eff} \hat\rho - \hat\rho \hat {H}_\text{eff}\dg \big) + {\cal J}[\hat L_1] \hat\rho + {\cal J}[\hat L_2] \hat\rho,
\end{align}
The anti-Hermitian, effective Hamiltonian 
$\hat{H}_\text{eff} = -\frac{i}{2}(\hat{L}_1\dg \hat{L}_1 + \hat{L}_2\dg \hat{L}_2)$
has spectrum $-\frac{i \gamma}{2}( 0,2,1,1 )$ with associated eigenstates
\bse \label{evec:nonHH}
\bqa  
 \ket{{\rm w}_1} 
&=& 
\ket{\Phi^-(\epsilon)}, \\
\ket{{\rm w}_2} &=& \frac{1}{\sqrt{2+\epsilon}} \Big(\sqrt{1+\epsilon}\ket{\er\er}+\ket{\gc\gc}\Big), \\
\ket{{\rm w}_3} &=& f \ket{\er \gc} + g^* \ket{\gc \er}, \\
\ket{{\rm w}_4} &=& g \ket{\er \gc} - f^* \ket{\gc \er},
\eqa 
\ese 
where $f,g$ can be chosen arbitrarily as long as $|f|^2 + |g|^2 = 1$. The four eigenstates are mutually orthogonal (thus comprising a basis) and will be useful in the analyses below. Note that the first two are highly entangled for $\epsilon \ll 1$.

In the long-time limit, $t\rightarrow \infty$, the two-atom steady state, $\hat \rho_{\rm ss}$ of the master equation for $\epsilon \neq 0$ is the near-Bell state $\ket{\Phi^{-}(\epsilon)} = \ket{\text{w}_1}$~\cite{DarGil18}. The two-atom steady-state changes abruptly when the bath qubits are prepared in the maximally entangled Bell state $\ket{\Phi_B^{+}}$ ($\epsilon = 0$). In this case, the two Lindblad operators become Hermitian conjugates of one another, $\hat L_1 = \hat{L}^\dagger_2 = \sqrt{\gamma / 2} (\hat{\sigma}_{1} + \hat{\sigma}_{2}^\dagger) $, and the master equation in \erf{ME:Bell:phi} does not have a rank-1 steady state. Instead, $\hat \rho_{\rm ss}$ is in general a mixed state with support in both Bell subspaces and is determined by the initial state of the atoms $\hat \rho_0$. An exact form for $\hat \rho_{\rm ss}$ and an extensive study can be found in \crf{DarGil18}. We highlight two interesting cases here. First, consider an initial two-atom state $\ket{\Phi^-}$. This state does not evolve (and thus it is a steady state), in analogy to $\ket{\Phi^-(\epsilon)}$ for the near-Bell master equation discussed above. Second, consider an arbitrary state (pure or mixed) $\hat{\rho}_\perp^{\Phi^-}$ in the 3-dimensional subspace orthogonal to $\ket{\Phi^-}$, \emph{i.e.} $\bra{\Phi^-}\hat{\rho}_\perp^{\Phi^-}\ket{\Phi^-} = 0$. In this case, the master-equation steady state is
    \begin{align} \label{thesubspacestate}
        \hat{\rho}_\perp^{\Phi^-} &\longrightarrow \hat \rho_{\rm ss} = \frac{1}{3}\big(\op{\er\gc}{\er\gc} + \op{\gc\er}{\gc\er} + \op{\Phi^+}{\Phi^+}\big), 
    \end{align}
which is the fully mixed state in that subspace. Note that the final states in the mixture are eigenstates of $\hat{H}_\text{eff}$,  \erf{evec:nonHH}. 

\begin{figure*}
\centering
\includegraphics[scale=0.43]{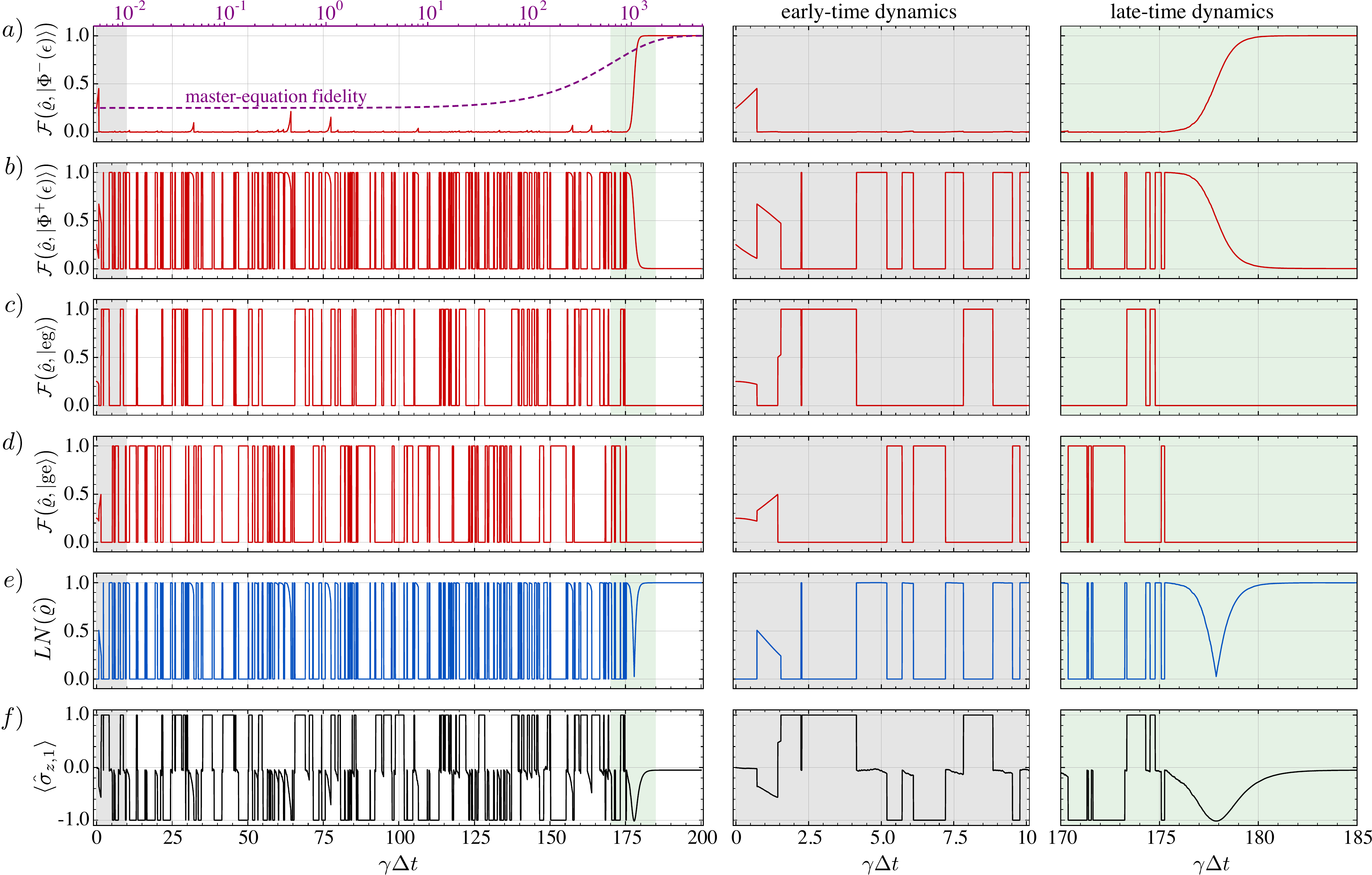}
\caption{\label{fig:loc:nmax} (Color online). A single, example quantum trajectory for a two-atom system interacting with a bath of qubits prepared in a near-Bell state, \erf{phi:pm:ep}, with $\epsilon = 0.1$. The atoms are prepared in the maximally mixed state.
a) Fidelity of the conditional two-atom state with the atomic steady state $\hat{\rho}_\text{ss} = \op{\Phi^-(\epsilon)}{\Phi^-(\epsilon)}$. For reference, the  fidelity is also shown for the average, master-equation dynamics (note the time scale is significantly different).
b-d) Fidelity of the conditional state with the three other eigenstates of the effective Hamiltonian, \erf{evec:nonHH}, with $f = g= 1$ giving $\ket{\text{w}_3} = \ket{\er \gc}$ and $\ket{\text{w}_4} = \ket{\gc \er}$.
e) Conditional-state entanglement as measured by the logarithmic negativity, \erf{log:neg}, which ultimately reaches the steady-state value $\approx 0.9992$, see \erf{LNss}.
f) Expectation value $\expt{ \hat{\sigma}_{z,1}}$. Note the quasi-diffusive behavior of the curve when it is not $\pm 1$.
}
\end{figure*}

\subsection{Local energy-basis measurements} \label{prob:2qbath:loc}

The procedure to find the conditional system state for each outcome is presented in \srf{loc:basis:theory} for bath qubits described by \erf{qubitstatedecomp} and measurements in the local energy basis ${\bf m}$, \erf{MB:local}. The Kraus operators can be read off from their general forms above in \erf{eq:Krauspsi1nojump} and \erf{eq:Krauspsi1jump} and making the proper replacements for the system operators as the bath-qubit amplitudes. 
With these Kraus operators, one can derive conditional difference equations or simply use the general form in \erf{genconddiffs} with appropriate replacements.

For typical two-outcome jump-type trajectories, the counterpart to infrequent jumps are periods of smooth evolution given by a no-jump map whose associated outcome occurs with probability near 1. In the four-outcome situation we consider here, things work differently. 
The outcome probabilities, found from the general forms given in \erf{prob:loc:gen}, scale in the following way:
\begin{subequations} \label{prob:v:nmax}
\bqa 
\wp_{m_1} &\sim& \frac{1}{2+\epsilon} - {\cal O}(\gamma \Delta t), \label{prob:v1:nmax}\\
\wp_{m_2} &\sim& \frac{1+\epsilon}{2+\epsilon} -  {\cal O}(\gamma \Delta t), \label{prob:v2:nmax} \\
\wp_{m_3} , \wp_{m_4} &\sim& {\cal O}(\gamma \Delta t) \label{prob:v34:nmax} 
.
\eqa
\end{subequations}
The two jump outcomes, $m_3$ and $m_4$, occur only rarely with probability ${\cal O}(\gamma \Delta t)$. The two no-jump outcomes, $m_1$ and $m_2$, each occur with probability near $\frac{1}{2}$---\emph{i.e.} the ``typical'' smooth no-jump evolution has been split in two via the two (roughly) equally likely outcomes. 
This is typically the hallmark of diffusive dynamics, except that the Kraus operators, \erf{eq:Krauspsi1nojump}, do not contain terms proportional to $\sqrt{\g{} \Delta t}$---nor do the conditional maps, \erf{K1mapappendix}.
Rather, they have terms proportional to $\g{} \Delta t$ multiplying two types of operator \erf{drho:loc1:nmax} and \erf{drho:loc2:nmax} contain two types of operator. The first are single-atom projectors, $\hat{\sigma}_j \hat{\sigma}^\dagger_j = \op{\gc}{\gc}_j$ and $\hat{\sigma}^\dagger_j \hat{\sigma}_j = \op{\er}{\er}_j$. The second are bonafide two-atom transition that each have one nontrivial operation over the two atoms: $\hat{\sigma}_1 \hat{\sigma}_2 = \op{\gc \gc}{\er \er}$ and $\hat{\sigma}_1\dg \hat{\sigma}_2\dg = \op{\er \er}{\gc \gc}$. Together, these no-jump maps drive quasi-diffusive atomic dynamics until a jump occurs.
In \srf{sec:eng:base}, we show that coarse graining over the no-jump outcomes removes the two-outcome dependence, thereby averaging out the quasi-diffusive nature to produce smooth evolution characteristic of typical no-jump evolution in quantum optical settings.

In the long-time limit, $t\rightarrow \infty$, the steady state $\hat \rho_{\rm ss}$ of the master equation is the other near-Bell state $\ket{\Phi^{-}(\epsilon)}$~\cite{DarGil18}. Since the master equation can be obtained by an average over trajectories, a rank-1 steady-state subspace requires that the trajectory equations also have $\hat \rho_{\rm ss}$ as the stationary state. 
The atoms are in a stationary state of conditional evolution when the two following conditions are met: outcomes with nonzero probability have trivial maps (proportional to the identity), and any outcomes with nontrivial maps have probability zero. This is indeed the case for the steady state $\ket{\Phi^{-}(\epsilon)}$. Once it is reached, one can verify that the two jump outcomes no longer occur ($\wp_{m_3}=\wp_{m_4}=0$), and the no-jump maps [Eqs.~(\ref{K1mapappendix})] are trivial,
    \begin{align}
        \hat{K}_{m_1} \ket{\Phi^-(\epsilon)} \propto \ket{\Phi^-(\epsilon)},  \quad   \hat{K}_{m_2} \ket{\Phi^-(\epsilon)} \propto \ket{\Phi^-(\epsilon)}.
    \end{align}

We show a sample trajectory in \frf{fig:loc:nmax}.
Using the Uhlmann-Jozsa fidelity between mixed state $\hat{\rho}$ and pure state $\ket{\psi}$
\begin{align} \label{2q:fid}
	\mathcal{F}(\hat{\rho}, \ket{\psi} )
	\coloneqq 
	\bra{\psi} \hat{\rho} \ket{\psi}
	,
\end{align}
we plot the fidelity of the conditional state $\hat{\varrho}$ with the pure steady state $\hat{\rho}_{\rm ss} = \op{\Phi^{-}(\epsilon)}{\Phi^{-}(\epsilon)}$ as well as the fidelities with the other eigenstates of the effective Hamiltonian given in \erf{evec:nonHH}. Indeed, the atoms do eventually fall into the steady state. Before that happens, the atoms become periodically entangled and unentangled as quantum jumps occur. We use the logarithmic negativity~\cite{Ple05},
\beq \label{log:neg}
 	LN(\hat{\rho}) := {\rm log}_2 \(\tr{\sqrt{\hat{\rho}_{\rm PT}\dg \hat{\rho}_{\rm PT}}}\),
\eeq
to quantify two-atom entanglement, where $\hat{\rho}_{\rm PT}$ is the partial transpose of the two-atom state taken with respect to the atomic subsystem Hilbert spaces. The logarithmic negativity lies in $[0,1]$, with the lower limit characterizing separable states and the upper limit maximally entangled states. The logarithmic negativity of the steady state, 
    \begin{equation} \label{LNss}
        LN[\hat{\rho}_\text{ss}] = {\rm log}_2 \Big( 1 + 2 \frac{\sqrt{1+\epsilon}}{2+\epsilon} \Big)
    \end{equation}
is achieved periodically before steady state is reached, since it is identical for $\ket{\Phi^+}$. The trajectory illustrates that entanglement is generated and destroyed both by the jump and no-jump evolutions (although the former is much more pronounced).  

We identify three distinct periods in the conditional dynamics in  \frf{fig:loc:nmax}. The first is an initial period of purification, which occurs until $\approx 2 \gamma \Delta t$ for this trajectory. The second is a long period of jumps to and from the $\{ \ket{\er \gc }, \ket{ \gc \er } \}$ subspace. The final is a slower, quasi-diffusive approach to the steady state beginning from $\approx 175 \g{} \Delta t$ in this trajectory.
To understand what is happening, we find the action of the Kraus operators on an arbitrary pure state, see Appendix~\ref{appendix:twoatom} for details. When the state is found in one the two pure states $\ket{\er \gc }$ or $\ket{ \gc \er }$ (as indicated by $\mathcal{F} = 1$ for either of these states), it does not evolve until an outcome (either $m_3$ or $m_4$) signals a jump to $\ket{\Phi^+(\epsilon)}$:
\begin{subequations} \label{eq:jumpout}
\begin{align} 
\hat{K}_{m_3}\typeasuperscript \ket{ \er \gc }  
    & \propto \ket{\Phi^+(\epsilon)} , \\
\hat{K}_{m_4}\typeasuperscript \ket{\gc \er}  
    & \propto \ket{\Phi^+(\epsilon)}  .     
\end{align}
\end{subequations}
If the same outcome is detected immediately afterwards, the jump process is exactly reversed:
\begin{subequations}
\begin{align}
\hat{K}_{m_3}\typeasuperscript \ket{\Phi^+(\epsilon)}  
    & \propto \ket{\gc \er},
\\
\hat{K}_{m_4}\typeasuperscript \ket{\Phi^+(\epsilon)}  
    & \propto \ket{\er \gc}.
\end{align}
\end{subequations}
Note that there are no jumps \emph{between} the two states in \erf{eq:jumpout}. Thus each time a jump occurs, a significant amount of entanglement is either created or destroyed, clearly shown by drastic flips in the logarithmic negativity from 0 to $ \approx 1$ and back again.

However, if a jump does not occur immediately after the state is projected into $\ket{\Phi^+(\epsilon)}$, then quasi-diffusive evolution proceeds according to 
\begin{subequations}
\begin{align}
\hat{K}_{m_1}\typeasuperscript \ket{\Phi^+(\epsilon)}
    & \propto (1-\g{} \Delta t) \ket{\Phi^+(\epsilon)} + \g{} \Delta t \ket{\Phi^-(\epsilon)} ,
    \\
\hat{K}_{m_2}\typeasuperscript \ket{\Phi^+(\epsilon)}
    & \propto (1-\g{} \Delta t) \ket{\Phi^+(\epsilon)} - \g{} \Delta t \ket{\Phi^-(\epsilon)} .
\end{align}
\end{subequations}
As this continues, the portion of the state in $\ket{\Phi^+(\epsilon)}$ steadily decreases (albeit stochastically) with population being slowly transferred to $\ket{\Phi^-(\epsilon)}$. At any time, a jump may occur, which resets the whole process. However, if no jump occurs, then the state ultimately lands in the steady state, as in the final stage of evolution in \frf{fig:loc:nmax}.

\subsubsection{Exact Bell-state bath} \label{exact:Bell:bath}

When the bath qubits are in the exact Bell state $\ket{\Phi_B^+}$, the two Lindblad operators in the master equation become Hermitian conjugates of one another.  
The invariant subspace of the master equation, \erf{ME:Bell:phi}, expands from rank 1 (pure steady state) to rank 4, where a given steady state is dependent on the initial two-atom state~\cite{DarGil18}. 

 \begin{table}[b]
 \def\arraystretch{1.5}
\begin{tabular}{ c|cc|cc|cc|cc  }
 \hline
 \hline
  & $m_1$ & $\wp_1$ & $m_2$ & $\wp_2$  & $m_3$ & $\wp_3$ & $m_4$ & $\wp_4$  \\
 \hline
 $\ket{\Phi^-}$ &  $\ket{\Phi^-}$ & $\frac{1}{2}$ & $\ket{\Phi^-}$  &  $\frac{1}{2}$ & --- & 0 & --- & 0
 \\
 $\ket{\Phi^+}$ 
 & $ \ket{\Phi^{+}}$ 
 & $\frac{1 - 2\g{}\Delta t}{2}$ 
 & $ \ket{\Phi^{+}}$ 
 & $\frac{1-2\g{}\Delta t}{2}$   & $\ket{\gc \er}$ & $\Delta t$ & $\ket{\er \gc}$ & $ \g{} \Delta t$
 \\
 $\ket{\er \gc}$ & $\ket{\er \gc}$ & $\frac{1 - \g{}\Delta t}{2}$  &  $\ket{\er \gc}$ & $\frac{1- \g{}\Delta t}{2}$  &  $\ket{\Phi^+}$ & $\g{}\Delta t$ & --- & 0 
 \\
 $\ket{\gc \er}$ & $\ket{\gc \er}$ & $\frac{1- \g{}\Delta t}{2}$  &   $\ket{\gc \er}$ & $\frac{1- \g{}\Delta t}{2}$  &  --- & 0 & $\ket{\Phi^+}$ & $\g{}\Delta t$
 \\ 
  \hline
  \hline
\end{tabular}
\caption{\label{tableBell} Conditional output states given outcome $m_j$ and associated probabilities $\wp_j$ when the bath qubits are prepared in the exact Bell state $\ket{\Phi^+_B}$ and a single interaction followed by measurement of the bath qubits. Each row is one of four two-atom input states in the leftmost column, chosen because of their association with the master-equation steady states. }
\end{table}

\begin{figure}[t]
\includegraphics[scale=0.41]{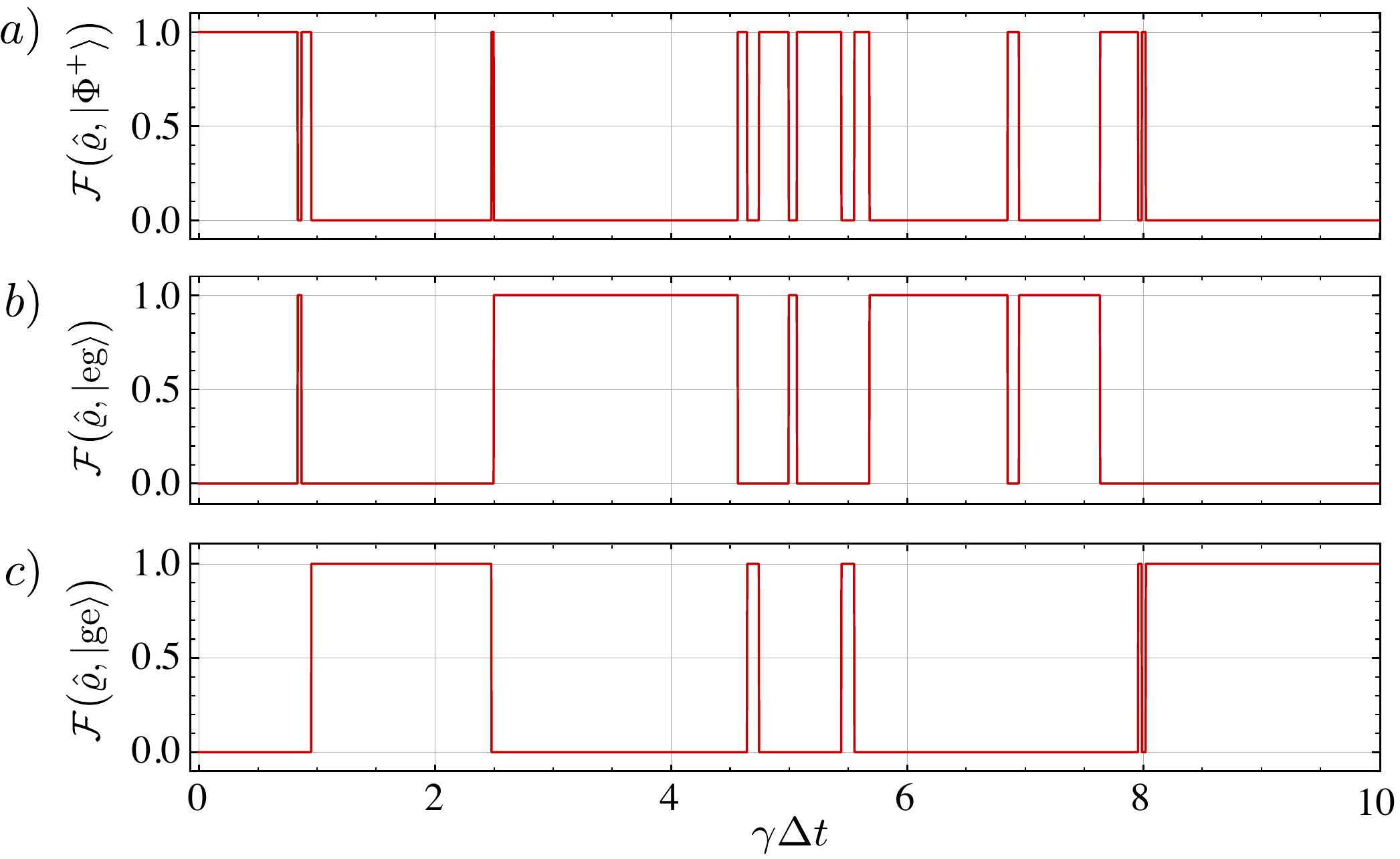}
\caption{\label{fig:fid_c_loc} (Color online). A single trajectory for the qubit bath in the exact Bell state $\ket{\Phi_B^+}$ and measurements in the local energy basis. Here, the two-atom initial state is $\ket{\Phi^+}$, and the resulting conditional state never leaves the subspace spanned by $\{ \ket{\Phi^+}, \ket{\er \gc}, \ket{\gc \er} \}$; fidelities with these states are shown, and the fidelity with $\ket{\Phi^-}$ is always zero.
An average over trajectories yields the state $\hat{\rho}^{\Phi^-}_\perp$, \erf{thesubspacestate}.
}
\end{figure}

The trajectory evolution is distinct from the average, master-equation evolution. We narrow our focus on the conditional evolution of the states in the set, $\{\ket{\Phi^-}, \ket{\Phi^+}, \ket{\gc \er}, \ket{\er \gc} \}$, as they play a crucial role in the long-time behavior of the master equation, \erf{thesubspacestate}---also see the discussion around \erf{eval:nonHH:genero}. We apply each of the four Kraus operators to these states to find the output state $\ket{\psi^{\rm out}} = \wp_j^{-1/2} \hat{K}_{m_j} \ket{\psi}$, and the probability of the outcome, $\wp_j$. A succinct summary of the results is given in Table~\ref{tableBell}; we discuss them in what follows.
The dynamical behavior of the Bell state $\ket{\Phi^-} $ is inherited from the case of near-Bell baths studied above, since
$\ket{\Phi^-(\epsilon=0)} = \ket{\Phi^-}$.
That is, if the conditional two-atom state reaches $\ket{\Phi^-}$, it does not experience further evolution at all regardless of outcome. 
The other three states, $\ket{\Phi^+}$, $ \ket{\gc \er}$, and $\ket{\er \gc}$ experience nontrivial evolution. However, none of these states evolves under the no-jump equations, as their evolution is governed entirely by jumps, which are signalled by outcomes $m_3$ and $m_4$. 
Atoms in the state $\ket{\Phi^+}$ periodically jump to $ \ket{\gc \er}$ and $\ket{\er \gc}$ and then jump back to $\ket{\Phi^+}$ eventually. 
However, the state never jumps directly between states in the $\{ \ket{\er \gc}, \ket{\gc \er} \}$ subspace. 
Thus, entanglement is created or destroyed during each jump. If the two-atom system begins in the $\{ \ket{\Phi^+}$, $ \ket{\gc \er}, \ket{\er \gc}\}$ subspace, it will remains trapped there, forever jumping between the three states as illustrated in \frf{fig:fid_c_loc}. An average over trajectories gives an equal mixture over these states, \erf{thesubspacestate}.

\subsection{Bell-state measurement basis} \label{Bellbasis_example}

As a counterpart to the local-basis measurements on for a two-atom system above, we consider the trajectories for the opposite case here: the entangled Bell-basis measurements described in Sec.~\ref{Bell:basis:theory}. From a practical perspective, this takes on a different physical setting than the situation above in that the bath qubits should (likely) be close together spatially in order to perform entangled measurements. 
We consider the same near-Bell bath-qubit state as above, $\ket{\Phi_B^+(\epsilon)}$ in \erf{Bell:phi:pm}, which means that the average over measurements gives the same master equation. The entangled Bell-state measurement basis still produces jump-type dynamics and respects the division into two no-jump maps and two jump maps. As before, the two no-jump maps together comprise quasi-diffusive dynamics. 

From the Kraus operators, \arf{appnA:joint}, it is straightforward to find the conditional maps and outcome probabilities for the quantum trajectories.
The outcome probabilities scale as
\bse \label{prob:u:nmax}
\begin{align}
\wp_{\Phi^\pm} &\sim \left(\frac{1}{2} \pm \frac{\sqrt{1+\epsilon}}{2+\epsilon}\right)-\left(\frac{1\pm\sqrt{1+\epsilon}}{\sqrt{2+\epsilon}}\right) {\cal O} (\gamma \Delta t), \label{prob:u1:nmax}\\
\wp_{\Psi^\pm} &\sim \left(\frac{1\pm\sqrt{1+\epsilon}}{\sqrt{2+\epsilon}}\right) {\cal O} (\gamma \Delta t). \label{prob:u34:nmax}
\end{align}
\ese
Unsurprisingly, the outcome corresponding to projection onto $\ket{\Phi^+}$ is the most likely (with a probability approaching one for small $\epsilon$) given that it has maximal overlap with the input bath state. The other no-jump outcome, $\ket{\Phi^-}$, has a small but nonvanishing probability until $\epsilon \rightarrow 0$, when it never occurs at all.

Generally, when the bath state coefficients are real-valued, such as we have here, the conditional difference equations in Appendix~\ref{Appendix:CDEBellbasis} can be written entirely in terms of the Lindblad operators. Under the condition that the probability of getting both no-jump outcomes is significantly greater than $\gamma \Delta t$, which occurs when
$\tfrac{1}{2} \pm (1+\epsilon)/\sqrt{2+\epsilon} > \gamma \Delta t$, they become,
\bse   \label{drho:joint:NBS}
\begin{align}
\Delta \vo_{\Phi^\pm} &= \Delta t f^\pm(\epsilon)\bigg\{ \sum_{\ell=1,2} \smallfrac{1}{2} {\cal M} \big[\hat L_\ell\dg \hat L_\ell \big]  \pm \nn \\
 &\hspace{45pt} \smallfrac{1}{2} {\cal M}\big[ \hat L_1 \hat L_2+\hat L_1\dg \hat L_2\dg \big]\bigg\}\vo , \label{drho:j1:qb}
\\
\Delta \vo_{\Psi^{\pm}} &=  {\cal G}[\hat{L}_\pm]\vo, \label{drho:j3:qb}
\end{align}
\ese
where $f^\pm(\epsilon) \coloneqq {\pm(2+\epsilon)}/{(1\pm(\sqrt{1+\epsilon})^2}$, and $\hat{L}_\pm$ are defined in \erf{eq:Lpm}.

A sample trajectory is shown in \frf{fig:joint:nmax} for near-Bell state bath qubits. After a period of jumps, the two-atom system reaches the master-equation steady state $\hat{\rho}_{\rm ss} = \op{\Phi^-(\epsilon)}{\Phi^-(\epsilon)}$, after which no more evolution occurs. This particular trajectory reaches $\hat{\rho}_{\rm ss}$ quickly compared to the average time $\gamma \Delta t \sim 10^3$---see \frf{fig:loc:nmax}(a). The reason why many trajectories take much longer than this is that the only path to the steady state is via no-jump dynamics from the state $\ket{\Phi^-(\epsilon)}$. Any jump will reset the process by forcing the state into the $\{ \ket{\Psi^\pm} \}$ subspace. The Bell-measurements differs significantly from jump dynamics when the bath qubits are monitored in the local energy basis. In that case, jumps project into the separable states $\ket{\er \gc}$ and $\ket{\gc \er}$.
Here, after the early-time dynamics, these jumps project the state into the specific maximally entangled Bell states $\ket{\Psi^\pm}$---logarithmic negativity is plotted as an entanglement proxy in \frf{fig:loc:nmax}(e). 
 
\blk

\begin{figure}
\includegraphics[scale=0.42]{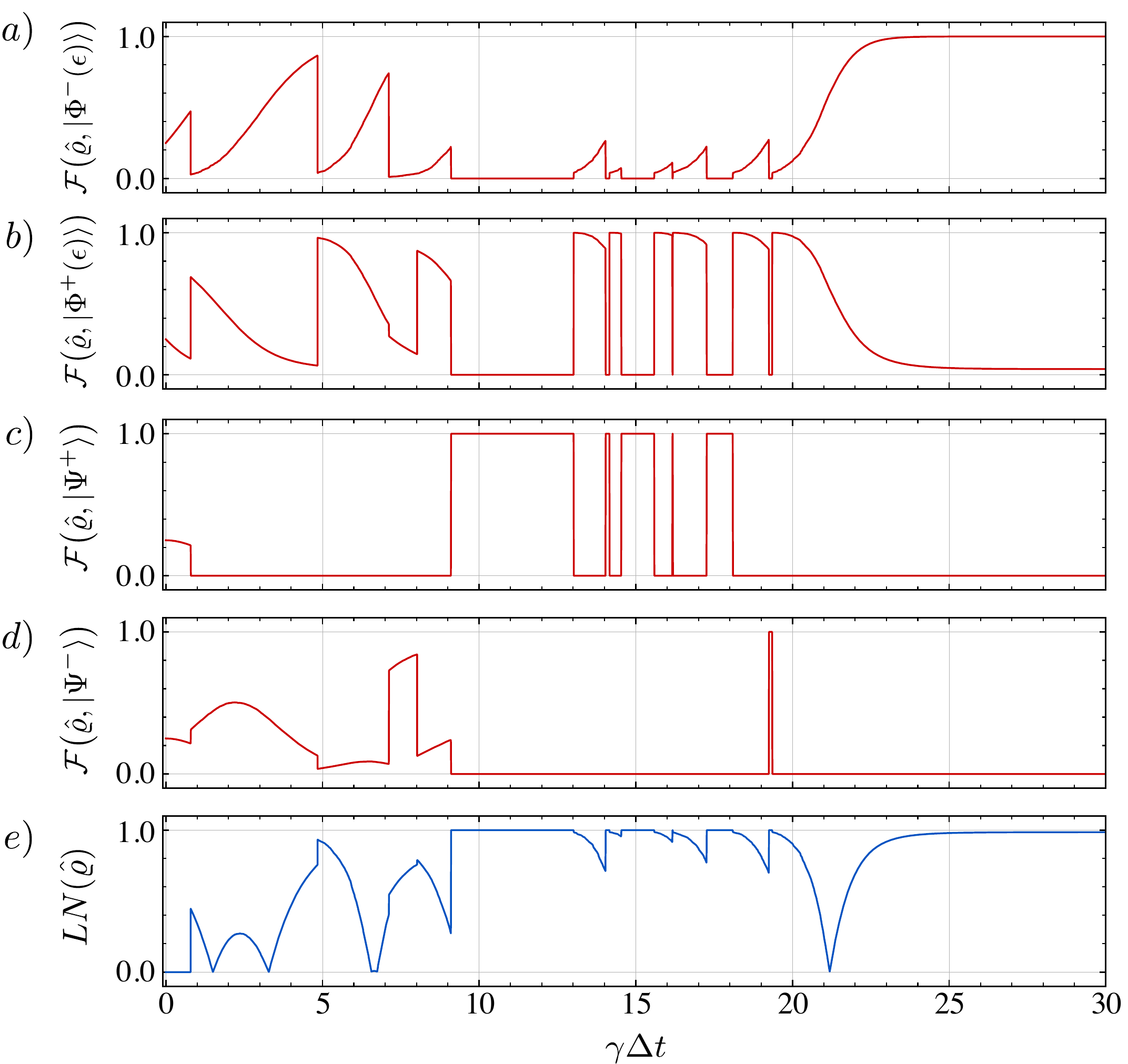}
\caption{\label{fig:joint:nmax} 
(Color online). A single, example quantum trajectory for a two-atom system interacting with a bath of qubits prepared in a near-Bell state, \erf{phi:pm:ep}, with $\epsilon = 0.1$ and measured in the maximally entangled Bell basis, \erf{MB:Bell}. 
a) Fidelity of the conditional two-atom state with the atomic steady state $\hat{\rho}_\text{ss}$ (see text).
b--d) Fidelity of the conditional state with the three other eigenstates of the effective Hamiltonian, \erf{evec:nonHH}, with $f = g = 1/\sqrt{2}$ giving $\ket{\text{w}_3} = \ket{\Psi^+}$ and $\ket{\text{w}_4} = \ket{\Psi^-}$.
e) Conditional-state entanglement as measured by the logarithmic negativity, \erf{log:neg}.
}
\end{figure}

\subsubsection{Exact Bell-state bath}
We now consider the exact Bell state bath mainly to point out some constrasting behavior compared to the local-energy basis measurements in \srf{exact:Bell:bath}. When the environment is initialized in the exact Bell state $\ket{\Phi^+}$, \erf{Bell:phi:pm}, 
the scaling of the outcome probabilities are obtained by setting $\epsilon=0$ in \erf{prob:u:nmax}:
\bse \label{prob:u:bell}
\bqa
\wp_{\Phi^+} &\sim& 1 - \mathcal{O}(\gamma \Delta t), \label{prob:u1:bell}\\
\wp_{\Phi^-} &\sim& \mathcal{O} \left[(\gamma \Delta t)^2\right] \approx 0, \label{prob:u2:bell}\\
\wp_{\Psi^\pm} &\sim&  \mathcal{O} (\gamma \Delta t). \label{prob:u34:bell}
\eqa
\ese
Because the measurement basis includes the input two-qubit state $\ket{\Phi^+}$, that outcome is much more likely than all the others. Also, the probability for outcome $\Phi^-$ is too high order in $\gamma \Delta t$, and thus it never occurs at all. This is in contrast to local energy-basis measurements, where both no-jump outcomes are equally likely, see \erf{prob:v:nmax} with $\epsilon = 0$.

For Bell measurements, the state never jumps directly between states in the $\{ \ket{\Psi^+}, \ket{\Psi^-}\}$ subspace. Instead, it jumps between from that subspace to the state $\ket{\Phi^+}$ and then potentially back again. In contrast to the local energy basis measurements, these jumps do not signal the creation or destruction of entanglement, as all of the three two-atom states are maximally entangled. 
However, for both measurement bases, if the two-atom system begins in the $\{ \ket{\Phi^+}$, $ \ket{\Psi^+}, \ket{\Psi^-}\}$ subspace, it will remain trapped there, forever jumping between the three states---as illustrated in \frf{fig:fid_c_joint}---and an average over trajectories gives an equal mixture over these states, \erf{thesubspacestate}.

\begin{figure}
\includegraphics[scale=0.41]{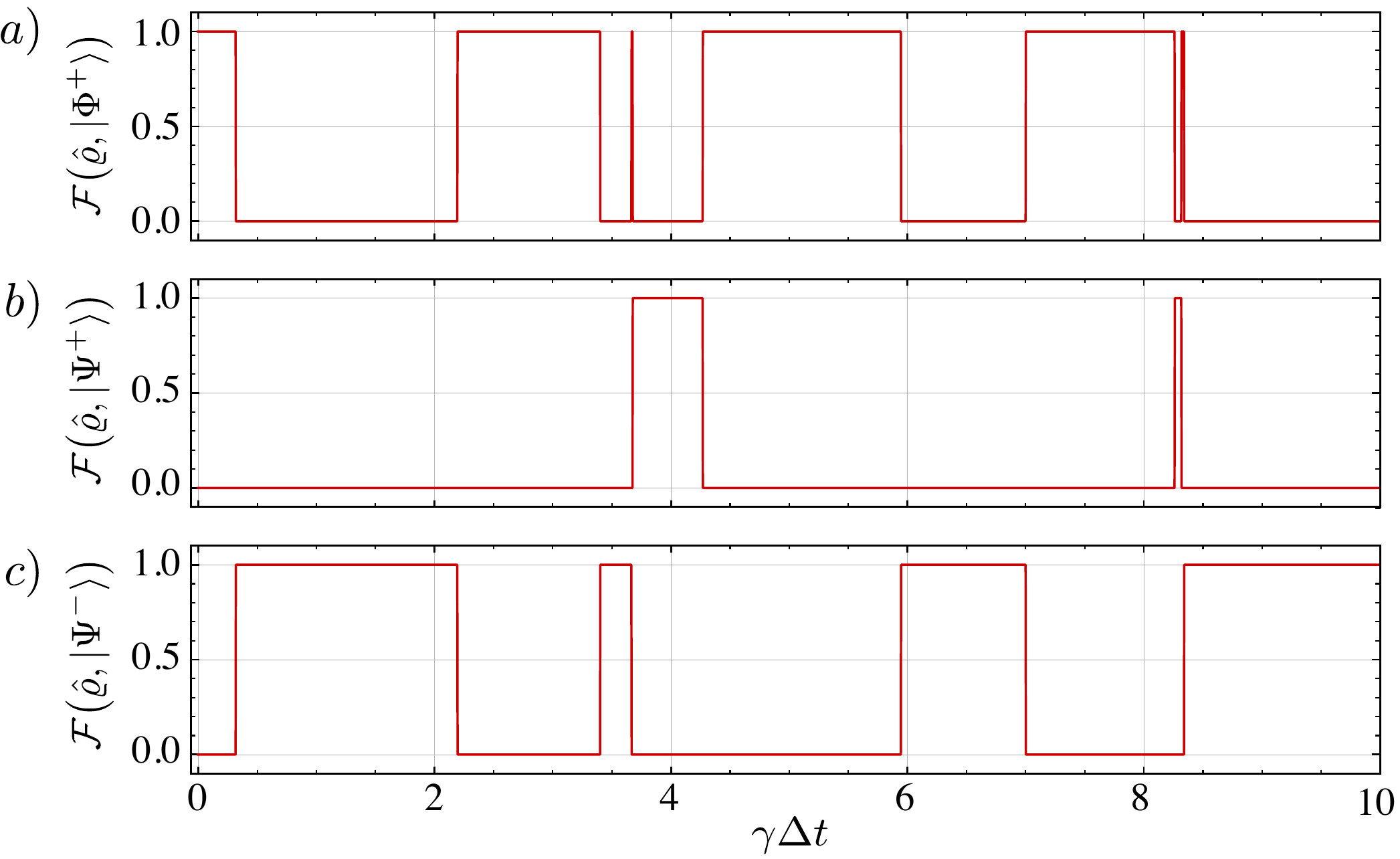}
\caption{\label{fig:fid_c_joint} (Color online).
(Color online). A single trajectory for the qubit bath in the exact Bell state $\ket{\Phi_B^+}$ and measurements in the Bell basis. Just as in \frf{fig:fid_c_loc}, the two-atom initial state is $\ket{\Phi^+}$, and the resulting conditional state never leaves the subspace spanned by $\{ \ket{\Phi^+}, \ket{\Psi^+}, \ket{\Psi^-} \}$; fidelities with these states are shown, and the fidelity with $\ket{\Phi^-}$ is always zero.
An average over trajectories yields the state $\hat{\rho}^{\Phi^-}_\perp$, \erf{thesubspacestate}.
}
\end{figure}

\section{Beyond two-qubit baths} \label{sec:beyond2qubits}
Thus far, we have focused our attention on two-qubit baths, as these types of bath contain essential features that distinguish them from the single-qubit baths typically considered.
A natural extension is to generalize to $n$-qubit environments, which was considered in the case of average master-equation dynamics in \crf{DarGil18}. Before moving to quantum trajectories, we present some of the points from the master-equation analysis.
Assuming all of the approximations made to derive the two-qubit ME for weak coupling, \erf{ME:gen}, a Markovian master equation can also be constructed for the $n$-qubit case~\cite{DarGil18}. 
Interestingly, the antidiagonal coherences in the $n>2$ qubit bath states do not enter into the ME at all~\cite{DarGil18}. 
This means that highly entangled Greenberger–Horne–Zeilinger (GHZ) state baths~\cite{GHZ07} (and the more general class of $X$-state baths where only diagonal and antidiagonal entries of $\hat \rho_B$ are nonzero~\cite{Eberly2012}) do not drive meaningful master-equation dynamics, \emph{i.e.} entanglement between subsystems will not be generated, only destroyed via simple decoherence channels. For example, a multi-atom system initialized in an $n$-qubit GHZ state \cite{GHZ07} would ultimately lose all of its entanglement~\cite{Wei10}. 

The absence of the antidiagonal bath coherences in the ME is directly connected to the weak-coupling limit. The unitary time evolution operator, \erf{Utaylor}, is truncated to second order in $\Delta t$ (recalling that $\lambda^2 \Delta t^2 = \gamma \Delta t$), which means that there is at most two-body operators acting on the bath's Hilbert space. Even for just three qubits in the bath, such operators can never transform $\ket{\er \er \er}$ to $\ket{\gc \gc \gc}$, so all system-bath terms will vanish under a bath trace if the initial bath state is the GHZ state
    \beq
        \ket{\psi_{B}} =  b_{\er\er\er} \ket{\er\er\er}+ b_{\gc\gc\gc} \ket{\gc \gc\gc} , \label{eq:3q:state}
    \eeq 
with $|b_{\er\er\er}|^2 + | b_{\gc\gc\gc}|^2 = 1$.
If one were to go beyond weak coupling in the interaction and include terms with many-body operators, that would allow further anti-diagonal coherences to contribute to master-equation dynamics.

We now move on to quantum trajectories in the context of three-qubit baths. At each time $\Delta t$, the bath qubits are initialized the GHZ state, \erf{eq:3q:state}, then they interact with the system via the weak-coupling unitary, after which they are measured in the local-energy basis via projections onto the set
\beq \label{measurment:basis:3q}
 {\bf q} = \{ \ket{\er\er\er}, \ket{\gc\gc\gc}, \ket{\er\er\gc}, \ket{\gc\gc\er}, \ket{\er\gc\er}, \ket{\gc\er\gc}, \ket{\er\gc\gc}, \ket{\gc\er\er} \},
\eeq 
with $q_j$ labeling each of the eight individual elements. The next steps for analyzing the system evolution require calculating the Kraus operators, conditional maps and the outcome probabilities. 
For this input state and measurement basis, there are two no-jump Kraus operators, \erf{Kraus:Opt:GHZnojump}, and six jump Kraus operators, \erf{Kraus:Opt:GHZjump}. This is due to the fact that, in general, elements of $\mathbf{q}$ that have overlap with the input state give no-jump Kraus operators, while those that have no overlap give jump Kraus operators. From the forms of these Kraus operators, it is evident that antidiagonal bath coherences, such as $b_{\er \er \er} b_{\gc \gc \gc}^*$, play no role in the conditional maps.
This is also evident in their absence in the scaling of the outcome probabilities,
\bse 
    \begin{align}
        \wp_{q_1} &\sim |b_{\er \er \er}|^2 \big[ 1 - \mathcal{O}(\gamma \Delta t) \big] \\
        \wp_{q_2} &\sim |b_{\gc \gc \gc}|^2 \big[ 1 - \mathcal{O}(\gamma \Delta t) \big] \\
        \wp_{q_3}, \wp_{q_5}, \wp_{q_8} &\sim |b_{\gc \gc \gc}|^2 \mathcal{O}(\gamma \Delta t) \\
        \wp_{q_4} , \wp_{q_6}, \wp_{q_7} &\sim |b_{\er \er \er}|^2 \mathcal{O}(\gamma \Delta t).
    \end{align}
    \ese 
As discussed above, their absence is tied to weak coupling. If one kept terms up to $(\lambda \Delta t)^3$ in the unitary evolution operator, the three-body bath operators that appear would introduce bath coherences into the conditional reduced-state dynamics. 

Since only a single bath-state amplitude, $b_{\er \er \er}$ or $b_{\gc \gc \gc}$, appears in each Kraus operator, all bath amplitudes cancel in the normalized conditional maps (although they still play a role in the probabilities). A result is that the conditional difference equations, 
\bse \label{diff:eqn:3q}
\begin{align}
    \Delta \vo_{q_1} &= -\smallfrac{1}{2}\gamma \Delta t \sum_\ell {\cal M} \big[\hat c_\ell \hat c_\ell\dg \big] \vo, \label{diff:eqn:3q:nj1} \\
    \Delta \vo_{q_2}  &= -\smallfrac{1}{2}\gamma \Delta t \sum_\ell {\cal M} \big[\hat c_\ell\dg \hat c_\ell \big] \vo, \label{diff:eqn:3q:nj2} \\
      \Delta \vo_{q_3} & = {\cal G}\big[\hat c_3\dg \big]\vo, \qquad 
      \Delta \vo_{q_4}  = {\cal G}[\hat c_3 ]\vo, \label{diff:eqn:3q:nj34} \\
      \Delta \vo_{q_5} & = {\cal G}\big[\hat c_2\dg \big]\vo, \qquad
      \Delta \vo_{q_6}  = {\cal G}[\hat c_2 ]\vo, \label{diff:eqn:3q:nj56} \\
      \Delta \vo_{q_8} & = {\cal G}\big[\hat c_1\dg \big]\vo, \qquad\Delta \vo_{q_7}  = {\cal G} [\hat c_1  ]\vo,
      \label{diff:eqn:3q:nj78}
\end{align}
\ese
which are valid when both $|b_{\er \er \er}|$ and $|b_{\gc \gc \gc}|$ are greater than $\sqrt{\gamma \Delta t}$,
contain no amplitudes at all.
Also, there are no coherent superpositions of system operators in the jump equations, compared to the two-qubit-bath setting in \erf{genconddiffs}. For system operators $\hat{c}_\ell$ act on separate subsystems, entanglement cannot be created, since the conditional evolution is entirely local regardless of the outcome. 

Trajectories for baths containing more qubits will exhibit similar behavior when the initial bath state is a GHZ-type state---the bath coherences will not affect the conditional system dynamics regardless of the outcome. However, this is not the only type of bath state. Others that have amplitudes differing by only a single excitation, for example $b_{\gc \gc \gc}$ and $b_{\er \gc \gc}$ for three qubits, can contribute.

\blk

\section{Conclusion} \label{sec:con}

In this work, we studied the stochastic conditional evolution of Markovian open quantum systems for repeated interactions with entangled two-qubit bath states.
We focused here on a particular type of two-qubit input state that could describe fully entangled Bell states. This produced both jump-type and diffusive-type conditional dynamics depending on the measurement basis. In a departure from typical quantum optical settings, more than one outcome can be associated with no-jump maps, giving rise to quasi-diffusive dynamics during evolution between jumps.
An interesting situation not considered here is where the bath state has support over the full two-qubit Hilbert space.
In this case, all the elements from any measurement basis have overlap with the input bath state, so no quantum jumps occur at all---only combinations of no-jump and diffusive dynamics.

Essential to the task of deriving quantum trajectories is the choice of measurement basis. 
A thorough analysis of the local-energy basis here set the stage for examining two others: local mixed-type and Bell-state bases. A general study of other bases has not been performed and could yield interesting results, with a focus on 
local bases, 
as they allow for more straightforward experimental implementations. If one is willing to relax the restriction to projective measurements, then more general POVMs for bath-qubit measurement can be considered. In addition, regardless of the measurement model, a thorough investigation of measurement statistics deserves its own study, as it could provide further understanding of novel effects such as quasi-diffusive trajectories.

The formalism here for two-qubit baths is mathematically equivalent to a single four-level qudit in the bath, although that setting has different physical interpretations. Consider a stream of identically prepared atoms sent through a microwave cavity, similar to that in Ref.~\cite{Haroche}, but where each atom is prepared in a superposition across a four-dimensional set of metastable Rydberg states. In this case, the coupling operators $\hat{c}_\ell$ and $\hat{c}^\dagger_\ell$ are all cavity operators (\emph{i.e.} they do act on separate subsystems). This could provide further tools for quantum control and state preparation of the cavity.

Various other avenues could be explored in future works. Notable is an extension beyond the weak-coupling regime, which has been done to various extents for single-qubit baths~\cite{GriMar16,Altamirano2017,Seah19beyond}. 
This would be particularly pertinent to baths with larger numbers of qubits, where many bath coherences whose effects are absent in weak coupling would play a role for stronger couplings. Another area of research is to devise quantum control protocols of the system which can make use of the measurement record to achieve certain goals. For instance, one may employ adaptive measurement techniques \cite{DarWis14,DarBra16} for entangling the subsystems at a faster rate.

Finally, there has recently been interest in using collisional models as a simulator for other systems of interest~\cite{CatGio21}, notably Gaussian~\cite{GroCom17} and few-excitation states of a single bosonic mode~\cite{Dabrowska2019}. Collisional models have the advantage that a sufficiently large quantum computer could execute these simulations. Multi-qubit environments may provide the strucuture to extend the scope of these simulations to multiple bosonic modes.

\blk

\acknowledgments
We thank Joshua Combes for fruitful discussions and Nicolas Menicucci for crucial help understanding the average over infinitesimal Kraus maps. This research was funded in part by the Australian Research Council Centre of Excellence for Engineered Quantum Systems (Project number CE170100009). B.Q.B. was additionally supported by the Australian Research Council Centre of Excellence for Quantum Computation and Communication Technology (Project No.\ CE170100012) and the Japan Science and Technology Agency through the MEXT Quantum Leap Flagship Program (MEXT Q-LEAP). S.D. was additionally  supported by the Engineering and Physical Sciences Research Council (EPSRC) the UK Quantum Technology Hub in Quantum Enhanced Imaging (QuantIC) EP/M01326X/1.

\newpage
	\appendix 
	\numberwithin{equation}{section}

\section{Mixed state baths} \label{appn:mixed:bath}

More generally, the environment can be in a mixed state $\hat{\rho}_B$. The conditional map on the system can be found by diagonalizing the environmental state,
	\begin{equation}
		\hat{\rho}_B = \sum_\ell p_\ell \op{\psi_{B_\ell}}{\psi_{B_\ell}}
		\, ,
	\end{equation}
where $0 \leq p_\ell \leq 1$ are mixture probabilities, and the pure states are orthogonal, $\ip{\psi_{B_\ell}}{\psi_{B_{\ell'}}} = \delta_{\ell,\ell'}$. The conditional map is then simply a mixture of maps of the form in Eq.~\eqref{map:condgen},
\bse 
\begin{align}   
{\mathcal E}_m(\vo) & \coloneqq \bra{m} \hat{U}_{SB} (\hat{\varrho} \otimes \hat{\rho}_B) \hat{U}^\dagger_{SB} \ket{m}, \\
 & = \sum_\ell p_\ell \hat K^{(\ell)}_m \vo \hat K_m^{(\ell)\dagger}, \label{map:condgenmix}
\end{align}
\ese 
where we have a set of Kraus operators for each state in the diagonalization, 
	\begin{equation} \label{eq:Krausgenmix}
		\hat K^{(\ell)}_m \coloneqq \bra{m} \hat U_{SB} \ket{\psi_{B_\ell}}
		\, .
	\end{equation}
The probability of outcome $m$ then becomes
	\begin{equation}\label{prob:condmix}
\wp_m = \sum_\ell p_\ell \text{Tr} \left[ \hat K^{(\ell)}_m \vo \hat K^{(\ell) \dagger}_m \right] = \text{Tr}\left[{\mathcal E}_m(\vo)\right].
	\end{equation}

\section{local energy-basis Kraus operators for bath qubits in the $\{\er \gc, \gc \er \}$ subspace} \label{appn_preA} 

Consider case where the bath state is $\ket{\psi_{B_\typeb}}$ in \erf{eq:state2}, \emph{i.e.} $p_\typea = 0$ and $p_\typeb = 1$, and the bath qubits are measured in the local energy basis.  The Kraus operators take on forms similar to those in Sec.~\ref{loc:basis:theory} depending on whether the outcomes are in the same Bell subspace as $\ket{\psi_{B_\typeb}}$---$\{\gc \er, \er \gc\}$ or the orthogonal one $\{\gc \gc,\er \er\}$:
\begin{subequations} \label{eq:KrausOpPsi:gen}
	\begin{align}
K_{m_1}^{(\typeb)}  &= -i \sqrt{\g{}\Delta t} \big(b_{\gc\er} \hat{c}_1 + b_{\er\gc} \hat{c}_2 \big) = -i \sqrt{ \Delta t}\, \hat{L}_3 , \\
K_{m_2}^{(\typeb)} &=  -i \sqrt{\g{}\Delta t}  \big(b_{\er\gc}  \hat{c}_1\dg + b_{\gc\er} \hat{c}_2\dg \big) = -i \sqrt{ \Delta t}\, \hat{L}_4 , \\
K_{m_3}^{(\typeb)} &= b_{\er\gc} \Big[ \hat{I}_S - \frac{\g{}\Delta t}{2}\big(   \hat{c}_1 \hat{c}_1\dg +  \hat{c}_2\dg \hat{c}_2 \big) \Big] - b_{\gc\er} \g{} \Delta t  \hat{c}_1 \hat{c}_2\dg, \\
K_{m_4}^{(\typeb)} &= b_{\gc\er} \Big[ \hat{I}_S - \frac{\g{}\Delta t}{2}\big(   \hat{c}_1\dg \hat{c}_1 + \hat{c}_2 \hat{c}_2\dg \big) \Big] - b_{\er\gc}\g{} \Delta t \hat{c}_1\dg \hat{c}_2.
\end{align}
\end{subequations}
To avoid confusion with Kraus operators in the main text, we include the additional superscript ${}^{(\typeb)}$ to indicate the input bath state.
The first two Kraus operators
 are ``jump'' operators, like \erf{eq:Krauspsi1jump}, and the second two 
 are ``no-jump'' operators, like \erf{eq:Krauspsi1nojump}.
These Kraus operators resolve the identity, $\sum_{j=1}^4 \hat K_{m_j}^{(\typeb)\dagger} \hat K_{m_j}^{(\typeb)} = \hat{I}_S $.

By substituting the corresponding Kraus operators into the \erf{map:cond}, we find the conditional maps for weak coupling (discarding terms higher order than $\gamma \Delta t$):
\begin{widetext}
\bse 
    \begin{align}
	\hat{K}_{m_1}^{(\typeb)} \hat \varrho \hat{K}^{(\typeb)\dagger}_{m_1} & =  \g{} \Delta t \big(  \abs{b_{\gc\er}}^2  \hat{c}_1 \vo \hat{c}_1\dg +   \abs{b_{\er\gc}}^2  \hat{c}_2 \vo \hat{c}_2\dg + b_{\er\gc} b_{\gc\er}^\ast  \hat{c}_2 \vo \hat{c}_1\dg + b_{\er\gc}^\ast b_{\gc\er}  \hat{c}_1 \vo \hat{c}_2\dg   \big) = \Delta t \, \hat L_3 \vo \hat L_3\dg , \\	
    \hat{K}_{m_2}^{(\typeb)} \hat \varrho \hat{K}^{(\typeb)\dagger}_{m_2} &= \g{} \Delta t \big( \abs{b_{\er \gc}}^2  \hat{c}_1\dg \vo \hat{c}_1 +  \abs{b_{\gc \er}}^2  \hat{c}_2\dg \vo \hat{c}_2 +  b_{\er\gc} b_{\gc\er}^\ast \hat{c}_1\dg \vo \hat{c}_2 +  b_{\er\gc}^\ast b_{\gc\er} \hat{c}_2\dg \vo \hat{c}_1  \big) = \Delta t \, \hat L_4 \vo \hat L_4\dg , \\
	\hat{K}_{m_3}^{(\typeb)} \hat \varrho \hat{K}^{(\typeb)\dagger}_{m_3} &=  \abs{b_{\er\gc}}^2 \vo - \g{} \Delta t \big(\smallfrac{1}{2}  \abs{b_{\er\gc}}^2 \hat{c}_1 \hat{c}_1\dg \vo + \smallfrac{1}{2}   \abs{b_{\er\gc}}^2 \hat{c}_2 \hat{c}_2\dg \vo 
	+ b_{\er\gc} b_{\gc\er}^\ast  \hat{c}\dg_1 \hat{c}_2 \vo
	+ {\rm H. c.}\big),\\	
	\hat{K}_{m_4}^{(\typeb)} \hat \varrho \hat{K}^{(\typeb)\dagger}_{m_4} &= \abs{b_{\gc\er}}^2 \vo - \g{} \Delta t\big(\smallfrac{1}{2}  \abs{b_{\gc\er}}^2  \hat{c}_1\dg \hat{c}_1 \vo + \smallfrac{1}{2} \abs{b_{\gc\er}}^2  \hat{c}_2\dg \hat{c}_2 \vo 
	+ b_{\er\gc} b^\ast_{\gc\er} \hat{c}_1\dg \hat{c}_2  \vo + {\rm H. c.}\big)  .
	\end{align}
	\ese 
The associated probabilities are
\bse 
\begin{align} 
\wp_{m_1}^{(\typeb)} &= \g{} \Delta t \big(  \abs{b_{\gc\er}}^2  \expt{\hat{c}_1\dg \hat{c}_1} +\abs{b_{\er\gc}}^2  \expt{\hat{c}_2\dg \hat{c}_2} + 2 \Re [ b_{\er\gc} b^*_{\gc\er}  \expt{\hat{c}_2\dg \hat{c}_1} ] \big) =\Delta t \, \expt{\hat L_3\dg \hat L_3},
\\
\wp_{m_2}^{(\typeb)} &= \g{} \Delta t  \big( \abs{b_{\er\gc}}^2 \expt{\hat{c}_1 \hat{c}_1\dg}  +  \abs{b_{\gc\er}}^2  \expt{\hat{c}_2 \hat{c}_2\dg} + 2 \Re [ b_{\er\gc} b^*_{\gc\er}  \expt{\hat{c}_2\dg \hat{c}_1} ] \big)
=\Delta t \, \expt{\hat L_4\dg \hat L_4},
\\
\wp_{m_3}^{(\typeb)} &=  \abs{b_{\er\gc}}^2 - \g{} \Delta t \big( \abs{b_{\er\gc}}^2  \expt{\hat{c}_1 \hat{c}_1\dg} +  \abs{b_{\er\gc}}^2  \expt{\hat{c}_2 \hat{c}_2\dg}  
+ 2 \Re [ b_{\er\gc} b^*_{\gc\er}  \expt{\hat{c}_2\dg \hat{c}_1} ] \big), 
\\
\wp_{m_4}^{(\typeb)} &= \abs{b_{\gc\er}}^2 -\g{} \Delta t \big( \abs{b_{\gc\er}}^2  \expt{\hat{c}_1\dg \hat{c}_1} +  \abs{b_{\gc\er}}^2 \expt{\hat{c}_2\dg \hat{c}_2} 2 \Re [ b_{\er\gc} b^*_{\gc\er}  \expt{\hat{c}_2\dg \hat{c}_1} ] \big).
\end{align}
\ese 
Note that the probabilities for jump outcomes here scale as 
$\wp_{m_1}, \wp_{m_2} \sim 1+\mathcal{O}(\gamma \Delta t)$, 
and the no-jump outcome probabilities scale as 
$\wp_{m_3} \sim \abs{b_{\er\gc}}^2 - \abs{b_{\er\gc}} \mathcal{O}(\gamma \Delta t)$ and $\wp_{m_4} \sim \abs{b_{\gc\er}}^2 - \abs{b_{\gc\er}}  \mathcal{O}( \gamma \Delta t)$.

\section{Conditional maps and outcome probabilities}  \label{appnA} 
	In this section, we present rather lengthy expressions for the conditional maps and outcome probabilities for the measurement bases described in the main text.
	
	\subsection{Local energy-basis measurements}   \label{appnA:loc}

The Kraus operators for the bath state $\ket{\psi_{B_\typea}}$, \erf{inputbathstate}, are given in the main text, \erfa{eq:Krauspsi1nojump}{eq:Krauspsi1jump}.
 By substituting the Kraus operators into the \erf{map:cond}, we find the conditional maps for weak coupling (to order $\gamma \Delta t$):
	\begin{subequations} \label{K1mapappendix} 
	\begin{align}
	\hat{K}_{m_1}\typeasuperscript \hat \varrho \hat{K}\typeasuperscriptdag_{m_1} 
& =  \abs{b_{\er\er}}^2 \vo - \g{} \Delta t \big(  \tfrac{1}{2} \abs{b_{\er\er}}^2 \hat{c}_1 \hat{c}_1\dg \vo + \tfrac{1}{2} \abs{b_{\er\er}}^2 \hat{c}_2 \hat{c}_2\dg \vo + 2 b_{\gc\gc} b_{\er\er}^\ast \, \hat{c}_1 \hat{c}_2 \vo + {\rm H. c.}\big) , \label{K1mapappendix1} 
\\
\hat{K}_{m_2}\typeasuperscript \hat \varrho \hat{K}\typeasuperscriptdag_{m_2} &= \abs{b_{\gc\gc}}^2 \vo - \g{} \Delta t \big( \tfrac{1}{2} \abs{b_{\gc\gc}}^2  \hat{c}_1\dg \hat{c}_1 \vo + \tfrac{1}{2} \abs{b_{\gc\gc}}^2  \hat{c}_2\dg \hat{c}_2 \vo + 2 b^\ast_{\gc\gc} b_{\er\er} \hat{c}_1^\dagger \hat{c}_2^\dagger  \vo + {\rm H. c.}\big) ,\label{K1mapappendix2}  \\
	\hat{K}_{m_3}\typeasuperscript \hat \varrho \hat{K}\typeasuperscriptdag_{m_3} &=  \g{} \Delta t \big( \abs{b_{\gc\gc}}^2  \hat{c}_1 \vo \hat{c}_1\dg +   \abs{b_{\er\er}}^2  \hat{c}_2\dg \vo \hat{c}_2 + b_{\gc\gc} b_{\er\er}^\ast \hat{c}_1 \vo \hat{c}_2 +  b_{\gc\gc}^\ast b_{\er\er} \hat{c}_2\dg \vo \hat{c}_1\dg  \big) = \Delta t\, \hat L_1 \vo \hat L_1\dg  , \\	
 	\hat{K}_{m_4}\typeasuperscript \hat \varrho \hat{K}\typeasuperscriptdag_{m_4} &= \g{} \Delta t \big( \abs{b_{\er\er}}^2  \hat{c}_1\dg \vo \hat{c}_1 +  \abs{b_{\gc\gc}}^2  \hat{c}_2 \vo \hat{c}_2\dg +  b_{\gc\gc} b_{\er\er}^\ast \hat{c}_2 \vo \hat{c}_1 +  b_{\er\er} b_{\gc\gc}^\ast \hat{c}_1\dg \vo \hat{c}_2\dg \big) = \Delta t\, \hat L_2 \vo \hat L_2\dg
 		.
	\end{align}
\end{subequations}
At the extremes, $\abs{b_{\er\er}} = 0$ or $\abs{b_{\er\er}} = 1$, one of the first two maps becomes trivial. For the former \erf{K1mapappendix1} is trivial and the probability of getting outcome $m_1$ vanishes, and for the latter, \erf{K1mapappendix2} is trivial, and the probability of getting outcome $m_2$ vanishes.

Tracing over the system gives the respective outcome probabilities, 
\bse 
\begin{align} \label{prob:loc:gen}
\wp_{m_1}\typeasuperscript 
&=  \abs{b_{\er\er}}^2 -  \g{} \Delta t \big( \abs{b_{\er\er}}^2 \expt{\hat{c}_1 \hat{c}_1\dg} +  \abs{b_{\er\er}}^2 \expt{\hat{c}_2 \hat{c}_2\dg}  + 2 \Re [ b_{\gc\gc} b_{\er\er}^* \expt{ \hat{c}_1 \hat{c}_2 }] \big), \\
\wp_{m_2}\typeasuperscript 
&= \abs{b_{\gc\gc}}^2 - \g{} \Delta t \big( \abs{b_{\gc\gc}}^2  \expt{\hat{c}_1\dg \hat{c}_1} +  \abs{b_{\gc\gc}}^2 \expt{\hat{c}_2\dg \hat{c}_2} + 2 \Re [ b_{\gc\gc} b_{\er\er}^* \expt{ \hat{c}_1 \hat{c}_2 }] \big), \\
\wp_{m_3}\typeasuperscript 
&= \g{} \Delta t \big( \abs{b_{\gc\gc}}^2  \expt{\hat{c}_1\dg \hat{c}_1} +  \abs{b_{\er\er}}^2  \expt{\hat{c}_2 \hat{c}_2\dg} + 2 \Re [ b_{\gc\gc} b_{\er\er}^* \expt{ \hat{c}_1 \hat{c}_2 }] \big) = \Delta t \, \expt{\hat L_1\dg \hat L_1} ,\\
\wp_{m_4}\typeasuperscript 
&=  \g{} \Delta t  \big( \abs{b_{\er\er}}^2 \expt{\hat{c}_1 \hat{c}_1\dg}  +  \abs{b_{\gc\gc}}^2  \expt{\hat{c}_2\dg \hat{c}_2} + 2 \Re [ b_{\gc\gc} b_{\er\er}^* \expt{ \hat{c}_1 \hat{c}_2 }] \big)=\Delta t \, \expt{\hat L_2\dg \hat L_2}
,
\end{align}
\ese 
where $\Re$ denotes the real part.

\subsection{Entangled Bell-basis measurements}  \label{appnA:joint}
It is straightforward to find explicit forms for these relations by employing  \erfa{eq:Krauspsi1nojump}{eq:Krauspsi1jump} given bath qubit state $\ket{\psi_{B_\typea}}$: 
\begin{subequations} \label{Appendix:bellKraus}
\begin{align}
    \hat K_{u_1}\typeasuperscript =& \frac{1}{\sqrt{2}} \Big\{ (b_{\er\er} + b_{\gc\gc}) \hat I_S - \frac{\g{} \Delta t}{2} \Big[\sum_\ell  \big( b_{\er\er} \hat{c}_\ell \hat{c}_\ell\dg + b_{\gc\gc} \hat{c}_\ell\dg \hat{c}_\ell\big) + 2 \big(b_{\gc\gc}\hat{c}_1 \hat{c}_2 + b_{\er\er}\hat{c}_2\dg \hat{c}_1\dg\big) \Big] \Big\}, 
    \\
    \hat K_{u_2}\typeasuperscript =& \frac{1}{\sqrt{2}} \Big\{ (b_{\er\er} - b_{\gc\gc}) \hat I_S - \frac{\g{} \Delta t}{2} \Big[\sum_\ell  \big( b_{\er\er} \hat{c}_\ell \hat{c}_\ell\dg - b_{\gc\gc} \hat{c}_\ell\dg \hat{c}_\ell\big) + 2 \big(b_{\gc\gc}\hat{c}_1 \hat{c}_2 - b_{\er\er}\hat{c}_2\dg \hat{c}_1\dg\big)\Big] \Big\}, 
    \\
    \hat K_{u_3}\typeasuperscript =& -i\sqrt{\frac{\g{} \Delta t}{2}} \Big[ b_{\gc\gc} \big(\hat{c}_1 + \hat{c}_2 \big) + b_{\er\er} \big( \hat{c}_1\dg + \hat{c}_2\dg \big)\Big]  
    = -i\sqrt{\frac{\Delta t}{2}} (\hat{L}_1 + \hat{L}_2) , 
    \\
    \hat K_{u_4}\typeasuperscript =& -i\sqrt{\frac{\g{} \Delta t}{2}} \Big[ b_{\gc\gc} \big(\hat{c}_1 - \hat{c}_2 \big) - b_{\er\er} \big( \hat{c}_1\dg - \hat{c}_2\dg \big)\Big]  
    = -i\sqrt{\frac{\Delta t}{2}} (\hat{L}_1 - \hat{L}_2), 
\end{align}
\end{subequations}
These Kraus operators produce jump-type dynamics just like the local energy-basis measurements above, with the first two yielding no-jump dynamics and the latter two jump dynamics. The conditional maps, outcome probabilities, and conditional difference equations can be found straightforwardly using the same techniques, and we omit their explicit forms here for brevity.

\subsection{Mixed local measurements: diffusive-type maps} \label{appn:xzBasis}

When the first bath qubit is measureed in the $\ket{\pm}=\frac{1}{\sqrt{2}}(\ket{\er}\pm \ket{\gc})$ basis, and the second in the $\ket{e},\ket{\gc}$ basis, the Kraus operators are
\bse 
   \begin{align} \label{jumpkraus}
        \hat{K}_{\pm\er}\typeasuperscript 
        &=
        \frac{1}{\sqrt{2}} \big( \hat{K}_{m_1}\typeasuperscript  \pm \hat{K}_{m_4}\typeasuperscript \big) 
        =
        \frac{1}{\sqrt{2}} \Big[ b_{\er\er} \Big( \hat{I}_S - \frac{ \g{}\Delta t}{2} \sum_{\ell=1,2} \hat{c}_\ell \hat{c}_\ell\dg \Big) - b_{\gc\gc}\g{} \Delta t  \hat{c}_1 \hat{c}_2 \mp i \sqrt{\g{} \Delta t} \big(b_{\er\er}\hat{c}_1\dg + b_{\gc\gc} \hat{c}_2 \big) \Big], 
        \\
        \hat{K}_{\pm\gc}\typeasuperscript 
        &= 
        \frac{1}{\sqrt{2}} \big( \hat{K}_{m_3}\typeasuperscript  \pm \hat{K}_{m_2}\typeasuperscript\big) 
        =
        \frac{1}{\sqrt{2}} \Big[\pm b_{\gc\gc} \Big( \hat{I}_S - \frac{\g{} \Delta t}{2} \sum_{\ell=1,2}  \hat{c}^\dagger_\ell \hat{c}_\ell \Big) {\mp} b_{\er\er} \g{} \Delta t  \hat{c}^\dagger_1 \hat{c}^\dagger_2 - i \sqrt{\g{} \Delta t}  \big(b_{\gc\gc} \hat{c}_1 + b_{\er\er} \hat{c}_2\dg \big)\Big], 
    \end{align}
    \ese 
where the Lindblad operators $\hat{L}_j$ are given in \erf{eq:jumpops}.
The corresponding conditional maps are
\bse \label{maps:XZ}
\begin{align}
        \hat{K}_{{\pm\er}}\vo \hat{K}_{{\pm\er}}^{\dagger} &=
        \frac{1}{2} \bigg\{ \abs{b_{\er\er}}^2 \vo - \gamma \Delta t 
        \bigg[ 
        \Big(\smallfrac{1}{2} \abs{b_{\er\er}}^2  \sum_{\ell=1,2} \vo \hat{c}_\ell \hat{c}_\ell\dg + b^*_{\gc\gc}  b_{\er\er} \vo \hat c_1\dg \hat c_2\dg + \text{H.c.} \Big)
        - \frac{1}{\gamma} \hat L_2 \vo \hat L_2\dg \bigg] 
        \pm i\sqrt{ \Delta t} \left( b_{\er\er} \vo \hat L_2\dg -  b^*_{\er\er} \hat L_2 \vo \right)  \bigg\}, 
        \\
        \hat{K}_{{\pm\gc}}\typeasuperscript\vo \hat{K}_{{\pm\gc}}^{\dagger} 
        &=
        \frac{1}{2} \bigg\{ \abs{b_{\gc\gc}}^2 \vo - \gamma \Delta t \bigg[ \Big(\smallfrac{1}{2} \abs{b_{\gc\gc}}^2  \sum_{\ell=1,2} \vo \hat{c}_\ell\dg \hat{c}_\ell + b_{\gc\gc}  b^*_{\er\er} \vo \hat c_1 \hat c_2 + \text{H.c.} \Big) - \frac{1}{\gamma}\hat L_1 \vo \hat L_1\dg \bigg] 
        \pm i\sqrt{ \Delta t} \big( \, b_{\gc\gc} \vo \hat L_1\dg -   b^*_{\gc\gc} \hat L_1\vo \big) \bigg\}, 
\end{align}  
\ese 
and the outcome probabilities are the following:
\bse
\begin{align}
\wp_{{\pm \er}} 
    &=
    \frac{1}{2} \bigg[ \abs{b_{\er\er}}^2 - \gamma \Delta t \Big(\abs{b_{\er\er}}^2  \expt{\hat{c}_2 \hat{c}_2\dg} - \abs{b_{\gc\gc}}^2 \expt{\hat{c}_2\dg \hat{c}_2}\Big) 
    \pm i\sqrt{ \Delta t} \expt{ b_{\er\er} \hat L_2\dg - b_{\er\er}^* \hat L_2} \bigg], 
    \\
    \wp_{{\pm \gc}} 
    &=
    \frac{1}{2} \bigg[ \abs{b_{\gc\gc}}^2 + \gamma \Delta t \Big(\abs{b_{\er\er}}^2  \expt{\hat{c}_2 \hat{c}_2\dg} - \abs{b_{\gc\gc}}^2 \expt{\hat{c}_2\dg \hat{c}_2}\Big) 
    \pm i\sqrt{ \Delta t} \expt{ b_{\gc\gc} \hat  L_1\dg - b_{\gc\gc}^* \hat L_1} \bigg].    
\end{align}
\ese

\blk
\section{Conditional difference equations} \label{appn_conddiffeqs}
Deriving the conditional difference equations is straightforward using the following procedure. At each time, use the Kraus operators to find the conditional state for each outcome, and then normalize that state by the associated measurement probability. The Kraus operators and probabilities for various measurement bases can be found in Appendix~\ref{appnA}.
Then, subtracting the system state at the previous time gives the conditional difference equations. 

We focus here on the local energy basis and Bell-state basis measurements. The conditional difference equations for mixed local basis measurements can be derived with the same methods.

\subsection{Local energy-basis measurements} \label{appendix:conddiffenergybasis}
The conditional maps and probabilities for local energy-basis measurements are given in Appendix~\ref{appnA:loc}. The two conditional difference equations for jumps can be written down immediately, but for the no-jump outcomes, $m_1$ and $m_2$, they can be simplified in the weak coupling limit with some work. For outcome $m_1$, the conditional difference equation is
\bse 
\begin{align}
    \Delta \vo_{m_1} = \frac{1}{\wp_{m_1}\typeasuperscript} \hat{K}_{m_1}\typeasuperscript \vo \hat{K}\typeasuperscriptdag_{m_1} - \vo 
    =\frac{|b_{\er \er}|^2 \vo - \frac{\gamma \Delta t}{2} |b_{\er \er}|^2 \big( \hat{c}_1 \hat{c}_1\dg \vo +  \hat{c}_2 \hat{c}_2\dg \vo + {\rm H. c.} \big) - \frac{\gamma \Delta t}{2} \big( b_{\gc \gc} b^*_{\er \er} \hat{c}_1 \hat{c}_2 \vo + {\rm H. c.}\big)}{|b_{\er \er}|^2 - \gamma \Delta t |b_{\er \er}|^2 \big(  \expt{\hat{c}_1 \hat{c}_1\dg} + \expt{\hat{c}_2 \hat{c}_2\dg} \big)  - \gamma \Delta t \, \Re[ b_{\gc \gc} b^*_{\er \er} \expt{\hat{c}_1 \hat{c}_2}]  } - \vo, 
\end{align}
\ese 
Assuming that $|b_{\er \er}|^2 > \gamma \Delta t$, 
\bse 
\begin{align}
    \Delta \vo_{m_1} 
    &=\frac{ \vo - \frac{\gamma \Delta t}{2} \big( \hat{c}_1 \hat{c}_1\dg \vo +  \hat{c}_2 \hat{c}_2\dg \vo - 2\frac{ b_{\gc \gc} }{b_{\er \er}} \hat{c}_1 \hat{c}_2 \vo + {\rm H. c.}\big)}{1 - \gamma \Delta t \big(  \expt{\hat{c}_1 \hat{c}_1\dg} + \expt{\hat{c}_2 \hat{c}_2\dg} - 2\Re \big[ \frac{ b_{\gc \gc} }{b_{\er \er}} \expt{\hat{c}_1 \hat{c}_2} \big] \big) } - \vo, 
    \\
    &\approx \left[\vo - \frac{\gamma \Delta t}{2} \bigg( \hat{c}_1 \hat{c}_1\dg \vo +  \hat{c}_2 \hat{c}_2\dg \vo - 2\frac{ b_{\gc \gc} }{b_{\er \er}} \hat{c}_1 \hat{c}_2 \vo + {\rm H. c.}\bigg) \right] \left[ 1 + \gamma \Delta t \bigg(  \expt{\hat{c}_1 \hat{c}_1\dg} + \expt{\hat{c}_2 \hat{c}_2\dg} - 2\Re \Big[ \frac{ b_{\gc \gc} }{b_{\er \er}} \expt{\hat{c}_1 \hat{c}_2} \Big] \bigg)  \right] - \vo, 
    \\
    &= \gamma \Delta t \bigg(  \expt{\hat{c}_1 \hat{c}_1\dg} + \expt{\hat{c}_2 \hat{c}_2\dg} - 2\Re \Big[ \frac{ b_{\gc \gc} }{b_{\er \er}} \expt{\hat{c}_1 \hat{c}_2} \Big] \bigg) \vo 
    - \gamma \Delta t \bigg( \frac{1}{2} \hat{c}_1 \hat{c}_1\dg \vo +  \frac{1}{2}  \hat{c}_2 \hat{c}_2\dg \vo - \frac{ b_{\gc \gc} }{b_{\er \er}} \hat{c}_1 \hat{c}_2 \vo + {\rm H. c.}\bigg) .
\end{align}
\ese 
In the third line we used that in weak-coupling $\gamma \Delta t \ll 1$ and expanded the denominator up to order $\gamma \Delta t$, discarding terms of higher order. 
Rearranging the terms and employing \erf{eq:measurementsuperoperator} leads to \erf{drho:loc1:nmax}. The equation for outcome $m_2$ follows in the same way.

\subsection{Bell-state basis} \label{Appendix:CDEBellbasis}

The Kraus operators for Bell-basis measurements are given in \erf{Appendix:bellKraus}. They can be used to find the conditional state and associated probabilities for each outcome. Togther, these can be used to write down the conditional difference equations. Under the conditions that $\frac{1}{2}|(b_{\er \er} \pm b_{\gc \gc})|^2 > \gamma \Delta t$, the no-jump equations may be re-expressed using a derviation similar to that in Sec.~\ref{appendix:conddiffenergybasis}, and the equations become,
\bse   \label{drho:joint:gen}
\bqa
\Delta \vo_{u_1} &=&  \frac{(\alpha_+ + \beta_+) \vo - \g{} \Delta t\left[ \sum_\ell \frac{1}{2} \big(\alpha_+ \vo \hat{c}_\ell \hat{c}_\ell\dg + \beta_+ \vo \hat{c}_\ell\dg\hat{c}_\ell + {\rm H.c.}\big) + \big(\alpha_+ \vo \hat{c}_1 \hat{c}_2 + \beta_+ \vo \hat{c}_1\dg \hat{c}_2\dg +{\rm H.c.}\big)\right]}{(\alpha_+ + \beta_+) - \g{} \Delta t \Big( \sum_{\ell} \expt{\Re[\alpha_+] \hat{c}_\ell \hat{c}_\ell\dg + \Re[\beta_+] \hat{c}_\ell\dg \hat{c}_\ell} + \expt{(\alpha_++\beta_+^*)\hat{c}_1 \hat{c}_2 + {\rm H.c.}} \Big) } -\vo , \label{drho:j1:gen}\\
\Delta \vo_{u_2} &=&  \frac{(\alpha_- - \beta_-)  \vo -  \g{}\Delta t\left[ \sum_\ell \frac{1}{2} \big(\alpha_- \vo \hat{c}_\ell \hat{c}_\ell\dg - \beta_- \vo \hat{c}_\ell\dg\hat{c}_\ell + {\rm H.c.}\big)  -  \big(\alpha_- \vo \hat{c}_1 \hat{c}_2 - \beta_- \vo \hat{c}_1\dg \hat{c}_2\dg +{\rm H.c.}\big)\right]}{ (\alpha_- - \beta_-) - \g{}\Delta t \Big( \sum_{\ell}  \expt{\Re[\alpha_-] \hat{c}_\ell \hat{c}_\ell\dg - \Re[\beta_-] \hat{c}_\ell\dg \hat{c}_\ell} - \,\expt{(\alpha_- -\beta_-^*)\hat{c}_1 \hat{c}_2 + {\rm H.c.}} \Big) } -\vo, \label{drho:j2:gen}\\
\Delta \vo_{u_3} &=& \frac{ \sum_\ell (b_{\gc\gc} \hat c_\ell + b_{\er\er} \hat c_\ell\dg)\vo (b_{\gc\gc} \hat c_\ell + b_{\er\er} \hat c_\ell\dg)\dg + \left[ (b_{\gc\gc} \hat c_1 + b_{\er\er} \hat c_1\dg)\vo(b_{\gc\gc} \hat c_2 + b_{\er\er} \hat c_2\dg)\dg+{\text H.c.}\right]}{\sum_\ell\expt{(b_{\gc\gc} \hat c_\ell + b_{\er\er} \hat c_\ell\dg)\dg(b_{\gc\gc} \hat c_\ell + b_{\er\er} \hat c_\ell\dg)}+\expt{ (b_{\gc\gc} \hat c_2 + b_{\er\er} \hat c_2\dg)\dg (b_{\gc\gc} \hat c_1 + b_{\er\er} \hat c_1\dg)+{\text H.c.}}} 
= {\cal G}\big[ \hat L_+ \big]\vo \label{drho:j3:gen}\\
\Delta \vo_{u_4} &=& \frac{ \sum_\ell (b_{\gc\gc} \hat c_\ell - b_{\er\er} \hat c_\ell\dg)\vo (b_{\gc\gc} \hat c_\ell - b_{\er\er} \hat c_\ell\dg)\dg - \left[ (b_{\gc\gc} \hat c_1 - b_{\er\er} \hat c_1\dg)\vo(b_{\gc\gc} \hat c_2 - b_{\er\er} \hat c_2\dg)\dg+{\text H.c.}\right]}{\sum_\ell\expt{(b_{\gc\gc} \hat c_\ell - b_{\er\er} \hat c_\ell\dg)\dg(b_{\gc\gc} \hat c_\ell - b_{\er\er} \hat c_\ell\dg)}-\expt{ (b_{\gc\gc} \hat c_2 - b_{\er\er} \hat c_2\dg)\dg (b_{\gc\gc} \hat c_1 - b_{\er\er} \hat c_1\dg)+{\text H.c.}}} 
= {\cal G}\big[ \hat L_- \big]\vo \label{drho:j4:gen}
\eqa
\ese
where 
$\alpha_ \pm = b_{\er\er}^*(b_{\er\er}\pm b_{\gc\gc})$ and
$\beta_ \pm = b_{\gc\gc}^*(b_{\er\er}\pm b_{\gc\gc})$,
and $\hat L_\pm$ are given in \erf{eq:Lpm}. 

\end{widetext}

\section{Conditional two-atom states} \label{appendix:twoatom}

In this appendix, we give various conditional two-atom states given bath state $\ket{\psi_B} = b_{\er \er} \ket{\er \er} + b_{\gc \gc} \ket{\gc \gc}$ and a pure input state for the atoms,
    \begin{equation} \label{genericinput}
        \ket{\psi} = c_0 \ket{\er\er} + c_1 \ket{\gc\gc} + c_2 \ket{\er\gc} + c_3 \ket{\gc\er}
        .
    \end{equation}
We focus here on local energy-basis measurements in \srf{loc:basis:theory}. Conditional states for other measurement bases can be constructed in a straightforward way by taking appropriate superpositions of the states presented here.

Since they act linearly, we apply each Kraus individually to the basis states in $\ket{\psi}$ above. We begin with the two no-jump outcomes, $\{ \er \er \}$ and $\{ \gc \gc \}$. The first has Kraus operator $\hat{K}_{m_1}\typeasuperscript$, which gives
\bse 
\begin{align} 
\hat{K}_{m_1}\typeasuperscript \ket{\er \er}
  & = b_{\er\er} \ket{\er \er} - b_{\gc\gc} \g{} \Delta t  \ket{\gc \gc},
  \\
\hat{K}_{m_1}\typeasuperscript \ket{\gc \gc}
  & = b_{\er\er} (1 - \g{} \Delta t) \ket{\gc \gc} ,
    \\
\hat{K}_{m_1}\typeasuperscript \ket{\er \gc}
  & = b_{\er\er} \big( 1 - \tfrac{ \g{} \Delta t}{2} \big) \ket{\er \gc}, 
    \\   
\hat{K}_{m_1}\typeasuperscript \ket{\gc \er}
  & = b_{\er\er} \big( 1 - \tfrac{ \g{} \Delta t}{2} \big)  \ket{\gc \er} .
    \end{align}
    \ese 
The second outcome $\{ \gc \gc \}$ gives
\bse 
\begin{align} 
\hat{K}_{m_2}\typeasuperscript \ket{\er \er}
  & = b_{\gc\gc} (1 - \g{} \Delta t) \ket{\er \er},
  \\
\hat{K}_{m_2}\typeasuperscript \ket{\gc \gc}
  & = b_{\gc \gc} \ket{\gc \gc} - b_{\er \er} \g{} \Delta t  \ket{\er \er}, 
    \\
\hat{K}_{m_2}\typeasuperscript \ket{\er \gc}
  & = b_{\gc\gc} \big( 1 - \tfrac{ \g{} \Delta t}{2} \big) \ket{\er \gc}, 
    \\   
\hat{K}_{m_2}\typeasuperscript \ket{\gc \er}
  & = b_{\gc\gc} \big( 1 - \tfrac{ \g{} \Delta t}{2} \big)  \ket{\gc \er} .
    \end{align}
    \ese 
Combining the above equations according to the coefficients in $\ket{\psi}$, gives the unnormalized conditional states,
\bse 
\begin{align} 
\hat{K}_{m_1}\typeasuperscript \ket{\psi}
    & = b_{\er\er} \ket{\psi} -\g{} \Delta t (c_0 b_{\gc \gc} + c_1 b_{\er\er}) \ket{\gc \gc} \nonumber \\
    & - \tfrac{1}{2}\g{} \Delta t \, b_{\er\er} (c_2 \ket{\er \gc} + c_3 \ket{\gc \er}) , \\
\hat{K}_{m_2}\typeasuperscript \ket{\psi}    
    & = b_{\gc\gc} \ket{\psi} -\g{} \Delta t (c_0 b_{\gc \gc} + c_1 b_{\er\er}) \ket{\er \er} \nonumber \\
    & - \tfrac{1}{2}\g{} \Delta t \, b_{\gc\gc} (c_2 \ket{\er \gc} + c_3 \ket{\gc \er}) , 
\end{align}
\ese 
We remark on a pathological case described previously in \srf{loc:basis:theory}. Consider the state $\hat{K}_{m_1}\typeasuperscript \ket{\psi}$ above. When $b_{\er \er} = 0$, the only surviving term is $\mathcal{O}(\g{} \Delta t)$. This means the probability of getting the outcome is of order $\wp_{m_1} \sim \mathcal{O}[(\g{} \Delta t)^2 ] = 0$, and $\hat{K}_{m_1}\typeasuperscript \ket{\psi}$ effectively vanishes. Alternatively, the density matrix element is $(\g{} \Delta t)^2$.
The same is true of $\hat{K}_{m_2}\typeasuperscript \ket{\psi}$ when $b_{\gc \gc} = 0$.

The two jump outcomes are $\{ \er \gc \}$ and $\{ \gc \er \}$. For outcome $\{ \er \gc \}$, we get
\bse 
    \begin{align} 
\hat{K}_{m_3}\typeasuperscript \ket{\er \er}
  & = -i \sqrt{\g{} \Delta t} \, b_{\gc\gc} \ket{\gc \er} ,
  \\
\hat{K}_{m_3}\typeasuperscript \ket{\gc \gc}
  & = -i \sqrt{\g{} \Delta t} \,  b_{\er\er}  \ket{\gc \er},
  \\
\hat{K}_{m_3}\typeasuperscript \ket{\er \gc}
  & = -i \sqrt{\g{} \Delta t} \, \big( b_{\gc\gc} \ket{\gc \gc} + b_{\er\er}  \ket{\er \er} \big),
    \\   
\hat{K}_{m_3}\typeasuperscript \ket{\gc \er}
  & = 0.
\end{align}
For outcome $\{ \gc \er \}$, we get
    \begin{align} 
\hat{K}_{m_4}\typeasuperscript \ket{\er \er}
  & = -i \sqrt{\g{} \Delta t} \, b_{\gc\gc} \ket{\er \gc}, 
  \\
\hat{K}_{m_4}\typeasuperscript \ket{\gc \gc}
  & = -i \sqrt{\g{} \Delta t} \,  b_{\er\er}  \ket{\er \gc},
  \\
\hat{K}_{m_4}\typeasuperscript \ket{\er \gc}
  & = 0,
    \\   
\hat{K}_{m_4}\typeasuperscript \ket{\gc \er}
  & = -i \sqrt{\g{} \Delta t} \, \big( b_{\gc\gc} \ket{\gc \gc} + b_{\er\er}  \ket{\er \er} \big).
\end{align}
\ese 
Combining these, we have the unnormalized conditional states
\bse 
\begin{align} 
\hat{K}_{m_3}\typeasuperscript \ket{\psi}
    & = -i \sqrt{\g{} \Delta t} \big[ (c_0 b_{\gc\gc} + c_1 b_{\er\er} ) \ket{\gc \er} \nonumber \\
     &+ c_2 ( b_{\gc\gc} \ket{\gc \gc} + b_{\er\er}  \ket{\er \er}) \big] ,
\\
\hat{K}_{m_4}\typeasuperscript \ket{\psi}
    & = -i \sqrt{\g{} \Delta t} \big[ (c_0 b_{\gc\gc} + c_1 b_{\er\er} ) \ket{\er \gc} \nonumber \\
     &+ c_3 ( b_{\gc\gc} \ket{\gc \gc} + b_{\er\er}  \ket{\er \er}) \big]. 
\end{align}
\ese 
Importantly, in the conditional state for $\{ \er \gc \}$, no term proportional to $c_3$ is present, because $\hat{K}_{m_3}\typeasuperscript \ket{\gc \er} = 0$. Further, the conditional state has no support on $\ket{\er \gc}$. When this outcome occurs, the state entirely leaves $\ket{\er \gc}$ with any support originally there, $c_3$, is transferred into the orthogonal Bell subspace. Similarly, in the conditional state for outcome $\{ \gc \er \}$, no term proportional to $c_2$ is present, because $\hat{K}_{m_4}\typeasuperscript \ket{\er \gc} = 0$. 

Using the above relations, one can write down useful forms for the system state immediately after a jump occurs in terms of the effective Hamiltonian eigenstates. An arbitrary two-atom state
\beq
\vo = \sum_{j,k \in \{\er, \gc \}}\sum_{j'\!,k'\in \{\er, \gc \}} \varrho_{jk,j'k'} \op{jk}{j'k'}.
\eeq
is transformed according to the following jump maps,
\bse \label{jump:map:gen:L1L2}
\begin{align}
    {\cal J}[\hat L_1] \hat  \varrho &= \gamma \Big[\eta_1 \op{\Psi_{B_{\Phi}}}{\Psi_{B_{\Phi}}} + \eta_2 \op{\gc\er}{\gc\er} + \nn\\
    &\hspace{20pt} \left( \eta_3 \op{\Psi_{B_{\Phi}}}{\gc\er} + \text{H.c.}\right) \Big], \\
    {\cal J}[\hat L_2] \hat  \varrho &= \gamma \Big[\mu_1 \op{\Psi_{B_{\Phi}}}{\Psi_{B_{\Phi}}} + \mu_2 \op{\er\gc}{\er\gc} + \nn\\
    &\hspace{20pt} \left( \mu_3 \op{\Psi_{B_{\Phi}}}{\er\gc} + \text{H.c.}\right) \Big],
\end{align}
\ese 
where the coefficients are determined by the matrix elements of the two-atom state as follows:
\bse
\begin{align}
    \eta_1 &=  \varrho_{\er\gc,\er\gc}, \\
    \eta_2 &=b_{\er\er}^*  (b_{\gc\gc} \varrho_{\er\er,\gc\gc}+b_{\er\er} \varrho_{\gc\gc,\gc\gc})  + \nn \\
    &\hspace{15pt} b_{\gc\gc}^* (b_{\er\er} \varrho_{\er\er,\gc\gc}^{*} + b_{\gc\gc} \varrho_{\er\er,\er\er}), \\
    \eta_3 &= b_{\er\er}^* \varrho_{\er\gc,\gc\gc} + b_{\gc\gc}^* \varrho_{\er\er,\er\gc}^{*}, \\
    \mu_1 &= \varrho_{\gc\er,\gc\er },  \\
    \mu_2 &= \eta_2, \\
    \mu_3 &= b_{\er\er}^*  \varrho_{\gc\er,\gc\gc} + b_{\gc\gc}^* \varrho_{\er\er,\gc\er}^{*}.
\end{align}
\ese 
\blk

\section{Three-qubit bath: Kraus operators and maps} \label{sec:appn:3q}

Given the three-qubit bath state,
    \beq
        \ket{\psi_{B}} =  b_{\er\er\er} \ket{\er\er\er}+ b_{\gc\gc\gc} \ket{\gc \gc\gc} , \label{eq:3q:stateAp}
    \eeq 
and measurements in the local energy basis given by projections onto the elements of $\mathbf{q}$, \erf{measurment:basis:3q}, we calculate the Kraus operators according to \erf{KrausWeak}. 
For this bath state and measurement basis, we have two no-jump Kraus operators
\bse \label{Kraus:Opt:GHZnojump}
\begin{align}
    \hat K_{q_1} &= b_{\er\er\er} \Big( \hat I_S - \frac{\gamma \Delta t}{2} \sum_{\ell=1}^3 \hat c_\ell \hat c_\ell\dg \Big), \\
    \hat K_{q_2} &= b_{\gc \gc\gc} \Big(\hat I_S - \frac{\gamma \Delta
    t}{2} \sum_{\ell=1}^3 \hat c_\ell\dg \hat c_\ell \Big), 
\end{align}
and six jump Kraus operators
\begin{align} \label{Kraus:Opt:GHZjump}
    \hat K_{q_3} &= -i \sqrt{\gamma \Delta t}\, b_{\er\er\er} \hat c_3\dg -\gamma \Delta t \,b_{\gc\gc\gc}\, \hat c_1 \hat c_2, \\
    \hat K_{q_4} &= -i \sqrt{\gamma \Delta t}\, b_{\gc\gc\gc} \hat c_3 -\gamma \Delta t \,b_{\er\er\er}\, \hat c_1\dg \hat c_2\dg, \\
    \hat K_{q_5} &= -i \sqrt{\gamma \Delta t}\, b_{\er\er\er} \hat c_2\dg -\gamma \Delta t \,b_{\gc\gc\gc}\, \hat c_1 \hat c_3, \\  
    \hat K_{q_6} &= -i \sqrt{\gamma \Delta t}\, b_{\gc\gc\gc} \hat c_2 -\gamma \Delta t \,b_{\er\er\er}\, \hat c_1\dg  \hat c_3\dg, \\ 
    \hat K_{q_7} &= -i \sqrt{\gamma \Delta t}\, b_{\gc\gc\gc} \hat c_1 -\gamma \Delta t \,b_{\er\er\er}\, \hat c_2\dg \hat c_3\dg, \\ 
    \hat K_{q_8} &= -i \sqrt{\gamma \Delta t}\, b_{\er\er\er} \hat c_1\dg -\gamma \Delta t \,b_{\gc\gc\gc}\, \hat c_2 \hat c_3.
\end{align}
\ese
In each of the jump Kraus operators, we retain the term proportional to $\gamma \Delta t$ simply to show the form. When used to calculate the maps, the Kraus operators are applied in pairs and these terms vanish as their effect is too high order in $\gamma \Delta t$.

Using these Kraus operators, it is straightforward to calculate the dynamical maps,
\bse
\begin{align}
    \hat K_{q_1} \vo \hat K_{q_1}\dg &= \abs{b_{\er\er\er}}^2 \bigg[ \vo -  \frac{\gamma \Delta t}{2} \sum_\ell \left(\vo \hat c_\ell \hat c_\ell\dg + \hat c_\ell \hat c_\ell\dg \vo\right)\bigg],\\
    \hat K_{q_2} \vo \hat K_{q_2}\dg &= \abs{b_{\gc \gc\gc}}^2 \bigg[ \vo -  \frac{\gamma \Delta t}{2} \sum_\ell \left(\vo \hat c_\ell\dg \hat c_\ell + \hat c_\ell\dg \hat c_\ell \vo\right)\bigg],\\ 
    \hat K_{q_3} \vo \hat K_{q_3}\dg &= \gamma \Delta t \abs{b_{\er\er\er}}^2 \hat c_3\dg \vo  \hat c_3, \\
    \hat K_{q_4} \vo \hat K_{q_4}\dg &= \gamma \Delta t \abs{b_{\gc \gc\gc}}^2 \hat c_3 \vo  \hat c_3\dg, \\
    \hat K_{q_5} \vo \hat K_{q_5}\dg &= \gamma \Delta t \abs{b_{\er\er\er}}^2 \hat c_2\dg \vo  \hat c_2, \\
    \hat K_{q_6} \vo \hat K_{q_6}\dg &= \gamma \Delta t \abs{b_{\gc \gc\gc}}^2 \hat c_2 \vo  \hat c_2\dg, \\
    \hat K_{q_7} \vo \hat K_{q_7}\dg &= \gamma \Delta t \abs{b_{\gc \gc\gc}}^2 \hat c_1 \vo  \hat c_1\dg, \\
    \hat K_{q_8} \vo \hat K_{q_8}\dg &= \gamma \Delta t \abs{b_{\er\er\er}}^2 \hat c_1\dg \vo  \hat c_1.
\end{align}
\ese 
The difference equations in \erf{diff:eqn:3q}
are obtained simply by normalizing the maps and thensubtracting the initial state.

To complete the analogy with the two-qubit baths considered in Ref.~\cite{DarGil18}, one could consider more general mixed states of the form
\begin{equation}
    \hat \rho_B = \sum_{n=1}^4 p_n \, \op{\psi_{B_n}}{\psi_{B_n}},
\end{equation}
which can also be represented as a block diagonal matrix just like two-qubit in \erf{rhoE:entgen}, making this an $X$-state. This state is a mixture of four mutually orthogonal pure states, $\ket{\psi_{B_n}}$,
each of which is in a subspace spanned by two GHZ states:
\bse
\begin{align}
\ket{\rm GHZ_1^\pm} &\coloneqq \frac{1}{\sqrt{2}} \big(\ket{\er\er\er} \pm \ket{\gc\gc\gc} \big), \\
\ket{\rm GHZ_2^\pm} &\coloneqq \frac{1}{\sqrt{2}} \big(\ket{\er\er\gc} \pm \ket{\gc\gc\er} \big), \\
\ket{\rm GHZ_3^\pm} &\coloneqq \frac{1}{\sqrt{2}} \big(\ket{\er\gc\er} \pm \ket{\gc\er\gc} \big), \\
\ket{\rm GHZ_4^\pm} &\coloneqq \frac{1}{\sqrt{2}} \big(\ket{\er\gc\gc} \pm \ket{\gc\er\er} \big),
\end{align}
\ese
That is, $\ket{\psi_{B_n}} \in {\rm span}\{\ket{\text{GHZ}_n^\pm}\}$. Finding the Kraus operators, maps, probabilities, and conditional difference equations for each pure state is a straightforward application of the techniques throughout this manuscript. The maps can then be statistically combined using the mixing probabilities $p_n$.


\bibliography{references}

\end{document}